\documentclass[fleqn,usenatbib]{mnras}

\usepackage{newtxtext,newtxmath}

\usepackage[T1]{fontenc}
\usepackage{ae,aecompl}

\usepackage{graphicx}	
\usepackage{amsmath}	
\usepackage{newtxtext,newtxmath}    
\usepackage{threeparttable, tablefootnote}

\DeclareMathOperator{\sech}{sech}

\title[Bulge, Disk, and Spirals in Galaxy Flybys]{Galaxy Flybys: Evolution of the Bulge, Disk, and Spiral Arms}

\author[A. Kumar et al.]{
Ankit Kumar,$^{1,2}$\thanks{E-mail: ankit4physics@gmail.com (AK)}
Mousumi Das,$^{1}$
and Sandeep Kumar Kataria$^{1,3}$
\\
$^{1}$Indian Institute of Astrophysics, Bengaluru, 560034, India\\
$^{2}$Joint Astronomy Program, Department of Physics, Indian Institute of Science, Bengaluru, 560012, India \\
$^{3}$School of Physics and Astronomy, Shanghai Jiao Tong University,No.800, Dongchuan Road, Minhang District, Shanghai, China. 
}

\date{Accepted XXX. Received YYY; in original form ZZZ}

\pubyear{2020}

\begin{document}
\label{firstpage}
\pagerange{\pageref{firstpage}--\pageref{lastpage}}
\maketitle

\begin{abstract}
Galaxy flybys are as common as mergers in low redshift universe and are important for galaxy evolution as they involve the exchange of significant amounts of mass and energy. In this study we investigate the effect of minor flybys on the bulges, disks, and spiral arms of Milky Way mass galaxies for two types of bulges -- classical bulges and boxy/peanut pseudobulges. Our N-body simulations comprise of two disk galaxies of mass ratios 10:1 and 5:1, where the disks of the galaxies lie in their orbital plane and the pericenter distance is varied. We performed photometric and kinematic bulge-disk decomposition at regular time steps and traced the evolution of the disk size, spiral structure, bulge sersic index, bulge mass, and bulge angular momentum. Our results show that the main effect on the disks is disk thickening, which is seen as the increase in the ratio of disk scale height to scale radius. The strength of the spiral structure $A_{2}/A_{0}$ shows small oscillations about the mean time-varying amplitude in the pseudobulge host galaxies. The flyby has no significant effect on non-rotating classical bulge, which shows that these bulges are extremely stable in galaxy interactions. However, the pseudobulges become dynamically hotter in flybys indicating that flybys may play an important role in accelerating the rate of secular evolution in disk galaxies. This effect on pseudobulges is a result of their rotating nature as part of the bar. Also, flybys do not affect the time and strength of bar buckling.

\end{abstract}




\section{Introduction}
\label{sec:intro} 
There are basically two types of galaxy interactions; mergers and flybys. It is well known that both phenomena play a major role in the hierarchical growth of the large scale structure \citep{Aarseth1979, Frenk1985}, as well as produce morphological changes in galaxies \citep{martin.etal.2018}. In the case of flybys, although the galaxies finally separate without merging, the interaction can have a strong effect on the galaxy disks \citep{arp.1966}. Flybys can be further categorized into major and minor interactions, depending upon the primary (host) to secondary (satellite) galaxy mass ratio. In general, if the ratio is greater than 3, it is classified as a minor interaction otherwise it is called a major interaction \citep{Stewart2009}. At early epochs, the universe was smaller in size and mergers were more frequent, hence they played an important role in the growth of galaxies \citep{Kauffmann1993, VanDenBergh1996, Murali2002, Stewart2008}. However, recent studies have shown that at later times, as the universe expanded, flybys became more common than mergers \citep{Sinha2012A}. Thus, although mergers may appear more important for changing galaxy morphology, the cumulative effect of flyby interactions may also be important since they are more frequent than mergers in the cosmic history of galaxies \citep{an.etal.2019}.

Early simulations clearly show the importance of flybys on the evolution of galaxies \citep{ToomreToomre1972, Barnes1992}. Later studies have found that flybys play a crucial role in producing tidally induced bars \citep{berentzen.etal.2004, lang.etal.2014, lokas.etal.2014}. The associated mass inflow and bar evolution can result in the formation of pseudobulges \citep{weinzirl.etal.2009} as well as kinematically decoupled cores \citep{derijcke.etal.2004}. The tidal interaction of flybys have an even stronger effect on the disks of the major galaxy, inducing the formation of spiral arms \citep{dubinski.etal.1999, renaud.etal.2015, Moreno.etal.2019} that result in strong star formation \citep{duc.etal.2018}. If the gas infall towards the center is large enough, then these effects can ultimately lead to starbursts \citep{mihos.hernquist.1994} and the triggering of active galactic nuclear (AGN) activity in galaxies \citep{combes.etal.2001}. Studies have shown that the tidal interaction can also result in the formation of warps in galaxy disks \citep{Kim2014}, as well as increase the stellar velocity dispersion, resulting in disk thickening \citep{Reshetnikov1997}. Other effects of flybys on stellar disks are the formation of stellar streams and tidal bridges \citep{duc.renaud.2013}. Studies suggest that flybys will also affect galaxy spin, although the effect is dependent on several factors such as orbit direction and the nature of the galaxies \citep{choi.yi.2017}.

In this study we investigate the effect of flybys on the two important stellar components of galaxies, their bulges and their disks. Bulges can be broadly classified into two types based on their morphology and the dynamics of their constituent stars; classical bulges that are approximately spherical in shape, and pseudobulges that appear disky and flatter in shape \citep{Kormendy2006, Drory2007, Athanassoula2008, Fisher2008A, Fisher2008B}. Classical bulges are dynamically hotter systems compared to pseudobulges that have more disky orbits \citep{Kormendy1993, Andredakis1994}. Classical bulges are thought to form from the monolithic collapse of gas clouds or clumps at early epochs, and then grow from the accretion of smaller galaxies \citep{Aguerri2001, Bournaud2005, Brooks2016}.

Pseudobulges can be further divided into two categories: disky pseudobulges and boxy/peanut pseudobulges. Disky pseudobulges are circular in shape but in the vertical direction they are as flat as the disks of their host galaxy. Hence, it is nearly impossible to detect them in edge-on galaxies. They are thought to form within the inner disks via star formation \citep{laurikainen.etal.2009}. Boxy/peanut bulges are more extended in the vertical direction and can hence be detected in edge-on galaxies. They are usually associated with bars in disk galaxies \citep{Friedli1990, Debattista2006, Gadotti2011}. Boxy/peanut pseudobulges, unlike classical bulges, are formed due to the secular evolution of galaxy disks, the buckling instability of bars \citep{Combes1981}, or via mergers with gas rich galaxies \citep{Keselman2012}. Using observations of galaxies in our local universe at distances less than 11 Mpc, \citet{Fisher2011} found that 80$\%$ of the galaxies with a stellar mass more than $10^{9} M_{\odot}$ contain either a pseudobulge or no bulge. This domination of pseudobulges in the local universe challenges the hierarchical models of galaxy formation, according to which the majority of galaxies should have classical bulges \citep{Frenk1985}.

In the literature, there have been several observational and numerical studies on the formation and evolution of bulges \citep{Kormendy2004, Athanassoula2005, Gadotti2009, Laurikainen2016}. It is clear that both secular processes and violent ones such as galaxy mergers are important for galaxy evolution \citep{tonini.etal.2016}. But most of the literature is biased towards only one kind of galaxy interaction -- galaxy mergers. The other kind of galaxy interaction, flybys, are largely ignored. This could be due to two reasons. The first is that the effect of flybys is lower than that of mergers. However, recent cosmological simulations have shown that galaxy flybys are as frequent as mergers in the low redshift universe \citep{Yee-Ellingson1995, Sinha2012A} and they may play an equally prominent role as mergers in the evolution of galaxies \citep{dimatteo.etal.2007, Cheng-Li2008, Kim2014, Sinha2015B}. The second reason is that bulges lie deep within the potential of a galaxy and the more massive ones make disks extremely stable against bar instability \citep{kataria.das.2018,K&D2020}. The spherical non-rotating classical bulge cannot be torqued effectively in flyby interactions unless there is some instability e.g. bar \citep{Kanak.etal.2012, Kanak.etal.2016}. Bars can be torqued in flyby interactions \citep{lokas.etal.2014} and so  boxy/peanut pseudobulges can also be torqued because they are essentially a part of the bar. The tidal torque of the satellite galaxies may thus change the kinematics of bars and their associated pseudobulges.

The tidal forces on galaxy disks due to flybys, are known to result in extended spiral arms. They can be similar to grand design spirals such as M51 \citep{dobbs.etal.2010}, or result in extremely large spiral arms such as in NGC 6872 \citep{horellou.koribalski.2007}. The interaction can lead to the formation of tidal tails, bridges and warped disks, but the overall morphology depends on many factors such as the pericenter distances, the galaxy orbits, and the relative masses of the galaxies \citep{oh.etal.2008}. Flybys may also affect the spin of galaxy disks, especially when the integrated effect of several interactions is included \citep{lee.etal.2018}. However, although the effect of tidal forces on galaxy disks has been explored in several earlier numerical studies \citep{walker.etal.1996}, it has not been quantified in terms of disk perturbation parameters, such as the Fourier components A$_2$/A$_0$, which is useful for measuring the effect of the tidal interaction, and is widely adopted in studies of bars \citep{kataria.das.2019}.

Thus, although there are several numerical studies of the tidal effect of flybys on galaxy disks, not much attention has been paid to its effect on the bulges or on quantifying the effects. For example, examining whether there are any changes in bulge sersic index or the strength of the spiral arms. In this paper we try to measure the changes in bulge and disk properties for different pericenter distances. We also include both classical and boxy pseudobulges in our models, disky pseudobulges will be studied later as their evolution possibly involves star formation. Although galaxy interactions do have a large parameter space, for simplicity in this study we focus on one of the parameters that causes the largest change in galaxy morphology -- the pericenter distance. 

In this study we have used a simple N-body approach; the effect of gas will be included in a future study. In the following sections we first discuss our simulation method which includes creating galaxy disks with classical and pseudobulges, describe how we have analysed the bulges and spiral arms, present our results and then discuss the implications of our work for observational studies.

\section{Simulations And Analysis}
In our simulations, we have fixed the mass ratio of the interacting galaxies at  5:1 and 10:1, as these ratio are similar to a wide range of minor flyby interactions in realistic scenarios. We have also restricted the disk angular momentum vectors to be perpendicular to the orbital plane of the interacting galaxies. The choice of parallel directions of the disk angular momenta of the major and minor galaxies was done to ensure maximum resonance (quasi-resonance) between the angular velocities of stars in the major galaxy and the peak angular velocity of the minor galaxy \citep{D'Onghia.etal.2010}.

\subsection{Galaxy Model: Halo and Disk}
\label{sec:model_holo_disk} 
We generated our model disk galaxies using the publicly available code GalIC \citep{Yurin2014}. Each galaxy has a live dark matter halo, and a stellar disk. In our models, the dark matter halo has a spherically symmetric Hernquist density distribution \citep{Hernquist1990},
\begin{equation}
    \rho_{dm}(r)=\frac{M_{dm}}{2\pi}\frac{a}{r(r+a)^3}
    \label{eqn:halo}
\end{equation}
where `$a$' is the scale radius of the dark matter halo and is related to the concentration parameter `$c$' of a corresponding NFW halo \citep{NFW1996} of mass $M_{dm}=M_{200}$ in the following manner,
\begin{equation}
    a=\frac{r_{200}}{c}\sqrt{2[\ln{(1+c)-\frac{c}{(1+c)}}]}
    \label{eqn:scale_radius}
\end{equation}
where $r_{200}$ is the virial radius of the galaxy (which is defined as the radius within which the mean density is 200 times the critical density of the universe) and $M_{200}$ is mass within the virial radius.

The stellar disk density has an exponential form in the radial direction and a $\sech^{2}$ form in the vertical direction
\begin{equation}
    \rho_{d}(R,z)=\frac{M_{d}}{4\pi z_{0} R_{s}^{2}}\exp(-{\frac{R}{R_{s}}}) \sech^{2}(\frac{z}{z_{0}})
    \label{eqn:disk}
\end{equation}
where $M_{d}$ is the total mass of the disk, $z_{0}$ is the scale height of the disk and $R_{s}$ is the scale radius of the disk.

\subsection{Galaxy Model: Bulge}
\label{sec:model_bulge}
To understand the effect of flyby events on the bulges and disks of galaxies, we have generated two types of bulges in the disk of the larger galaxy, which we call the major galaxy. The first is the classical bulge. Its density is derived from a spherically symmetric Hernquist potential \citep{Hernquist1990},
\begin{equation}
    \rho_{b}(r)=\frac{M_{b}}{2\pi}\frac{b}{r(r+b)^{3}}
    \label{eqn:bulge}
\end{equation}
where $M_{b}$ is the total mass of the bulge and `$b$' is the scale radius of the bulge. 

\begin{figure}
	\includegraphics[width=\columnwidth]{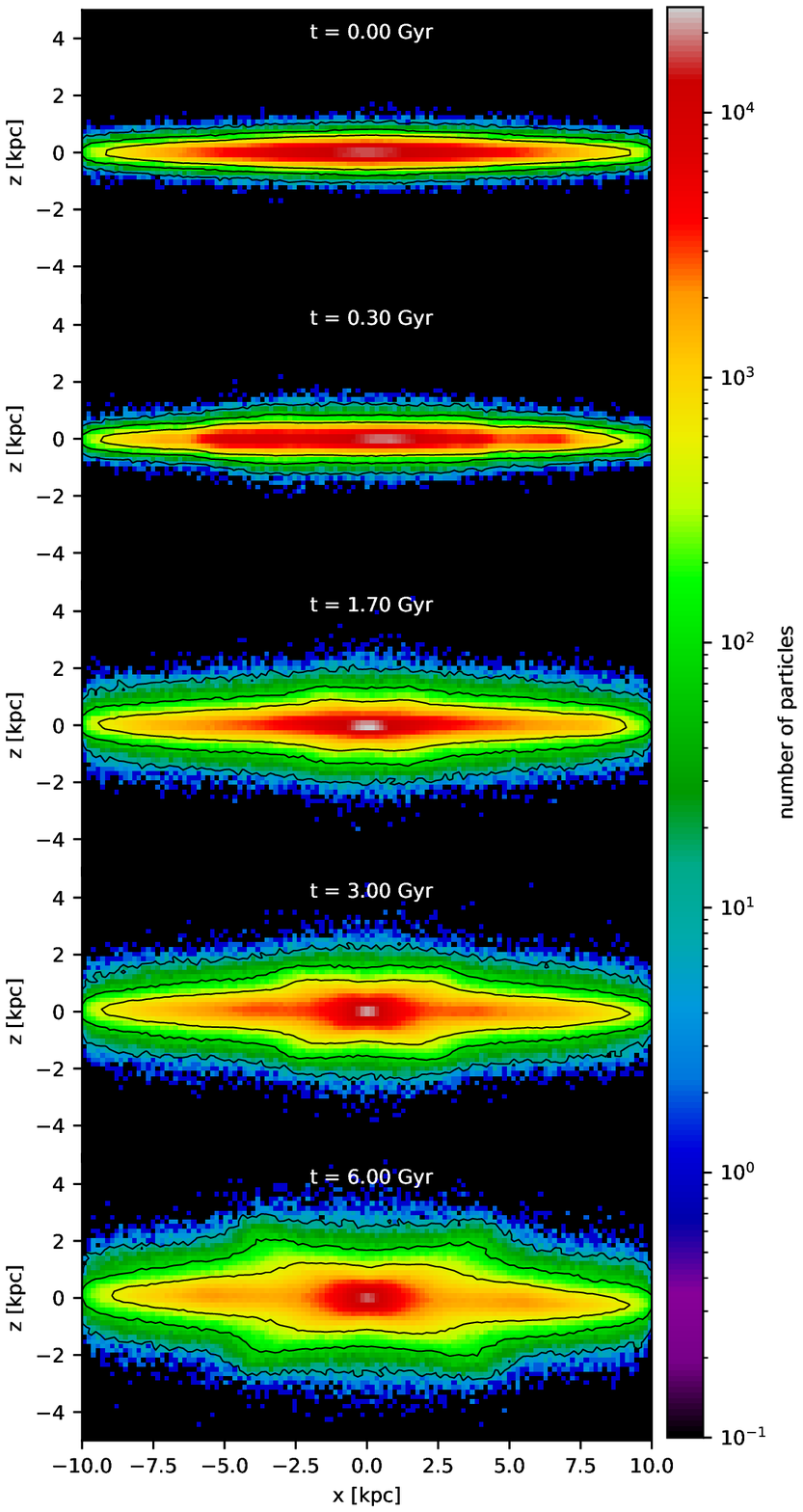}
    \caption{ The visual illustration of the pseudobulge formation. The top to bottom panels shows the time sequence of the edge-on view of the major galaxy, perpendicular to the bar, together with similar isodensity contours. From top to bottom, the five panels show the snapshots at t=0.0, 0.3 (1st peak in the bar strength), 1.7 (2nd peak in the bar strength), 3.0 (after the buckling), and 6.0 Gyr (beginning of the constant bar strength) respectively. The bottom panel displays the initial conditions for the flyby models of galaxies with a pseudobulge .}
    \label{fig:pb_formation}
\end{figure}

To obtain a pseudobulge in the disk of the major galaxy, we generated it naturally through the bar formation in a bulgeless galaxy. At a fixed disk scale height to scale radius ratio, we varied the angular momentum fraction of the disk to halo so that it can form a bar which can spontaneously buckle to form a pseudobulge (boxy/peanut bulge) after evolution. We evolved this model upto 10~Gyrs and calculated the bar strength using the ratio of m=2 to m=0 Fourier modes. The amplitudes of the $m$th Fourier mode at a radius $R$ is given by,
\begin{align}
    a_{m}(R) &= \sum_{i=1}^{N} m_{i}(R) \cos(m\phi_{i}), m=0,1,2,3, ... \\
    b_{m}(R) &= \sum_{i=1}^{N} m_{i}(R) \sin(m\phi_{i}), m=0,1,2,3, ... \\
    \intertext{where $m_{i}(R)$ and $\phi_{i}$ are the mass and the azimuthal angle of $i$th particle at radius $R$ respectively, and $N$ is the total number of particles at the same radial position. The strength of a bar is defined as}
    \frac{A_{2}}{A_{0}} &= max \Big[\frac{\sqrt{[a_{2}(R)]^2+[b_{2}(R)]^2}}{\sum_{i=1}^{N} m_{i}(R)} \Big]
    \label{eqn:bar_strenght}
\end{align}
\begin{figure}
	\includegraphics[width=\columnwidth]{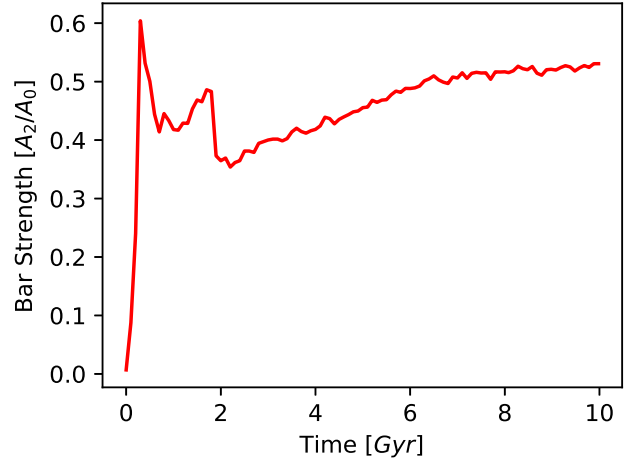}
    \caption{The evolution of bar strength in the pseudobulge model} until 10~Gyrs. Due to the buckling of the bar (the 2 peaks in the  $\frac{A_{2}}{A_{0}}$ value), the bar strength changes in the initial stage of evolution. After 6~Gyrs, it shows a nearly constant strength. This is where we take our initial model for a galaxy with a pseudobulge.
    \label{fig:bar_strength}
\end{figure}
Fig.~\ref{fig:bar_strength} shows the evolution of bar strength for the major galaxy with a pseudobulge ($PG_{PB}$ model) upto 10~Gyrs. The fluctuations in the bar strength at early epochs of the evolution is the indicator of bar buckling. The bar strength remains approximately constant after 6~Gyrs of evolution. This is where we take the initial conditions for our pseudobulge model i.e. we took the galaxy at 6~Gyr as the initial model for simulations of galaxy flybys with pseudobulges. The formation of the pseudobulge is graphically represented in fig.~\ref{fig:pb_formation} as a  time sequence of the galaxy evolution, using the edge-on view of the disk and bar.

\subsection{Choice of Model Parameters for Simulations}
\label{sec:sim_param}

\begin{table*}
\centering
\caption{Initial parameters of the primary (major) and secondary (minor) galaxies. In this table, $PG_{CB}$ = Primary galaxy with classical bulge, $PG_{PB}$ = Primary galaxy with pseudobulge, $SG_{5}$ = Secondary galaxy with 1/5 mass of primary galaxy, and $SG_{10}$ = Secondary galaxy with 1/10 mass of primary galaxy. (units $M_{\odot}$ = solar mass, $kpc$ = kiloparsec).}
\label{tab:initial_parameters}
\begin{threeparttable}
\begin{tabular}{lcccr}
\hline
Parameters & $PG_{CB}$ & $PG_{PB}$ & $SG_{5}$ & $SG_{10}$ \\
\hline
Total mass ($M$) & 1.2 $\times$ $10^{12} M_{\odot}$ & 1.2 $\times$ $10^{12} M_{\odot}$ & 2.4 $\times$ $10^{11} M_{\odot}$ & 1.2 $\times$ $10^{11} M_{\odot}$ \\
\hline
Halo spin parameter($\lambda$) & 0.035 & 0.035 & 0.035 & 0.035 \\
\hline
Halo concentration parameter($c$) & 10 & 10 & 11 & 11 \\
\hline
Disk mass fraction & 0.025 & 0.03 & 0.01 & 0.01 \\
\hline
Bulge mass fraction & 0.005 & 0.0 \tnote{a} & 0.002 & 0.002 \\
\hline
Disk spin fraction & 0.03 & 0.022 & 0.01 & 0.01 \\
\hline
Disk scale radius ($R_{s}$) & 3.80~kpc & 2.30~kpc & 1.95~kpc & 1.55~kpc \\
\hline
Disk scale height ($z_{0}$) & 0.38~kpc & 0.23~kpc & 0.195~kpc & 0.155~kpc \\
\hline
Halo particles($N_{Halo}$) & 2.5$\times 10^{6}$ & 2.5$\times 10^{6}$ & 5.0$\times 10^{5}$ & 2.5$\times 10^{5}$ \\
\hline
Disk particles($N_{Disk}$) & 1.5$\times 10^{6}$ & 2.5$\times 10^{6}$ & 3.0$\times 10^{5}$ & 1.5$\times 10^{5}$ \\
\hline
Bulge particles($N_{Bulge}$) & 1.0$\times 10^{6}$ & $0$ \tnote{a} & 2.0$\times 10^{5}$ & 1.0$\times 10^{5}$ \\
\hline
Total particles($N_{Total}$) & 5.0$\times 10^{6}$ & 5.0$\times 10^{6}$ & 1.0$\times 10^{6}$ & 5.0$\times 10^{5}$ \\
\hline
\end{tabular}
\begin{tablenotes}\footnotesize
\item [a] Initially this model does not have any bulge particles. A pseudobulge grows from the disk itself after a bar forms and buckles.
\end{tablenotes}
\end{threeparttable}
\end{table*}

Table~\ref{tab:initial_parameters} summarises the initial parameters of the primary (major) and secondary (minor) galaxies. The mass of the major galaxy is similar to that of the Milky Way galaxy in all the models, as suggested by many studies (see \cite{Wang.Haan2020} and reference therein). The halo spin parameter ($\lambda$) of the galaxies is chosen to be 0.035 which corresponds to the peak of the halo spin distribution given in \cite{Bullock2001}. The values of concentration parameter ($c$) lie well within the mass-concentration relation given by \cite{Wang.Bose2020}. The stellar mass fraction of the galaxies are chosen from the stellar to halo mass relation (SHMR) \citep{Behroozi2013, Moster2013}. The choice of disk scale length for the major galaxy is made such that the model $PG_{CB}$ does not show any bar feature after evolution but the model $PG_{PB}$ develops a bar which then buckles to form a pseudobulge.

\begin{figure}
	\includegraphics[width=\columnwidth]{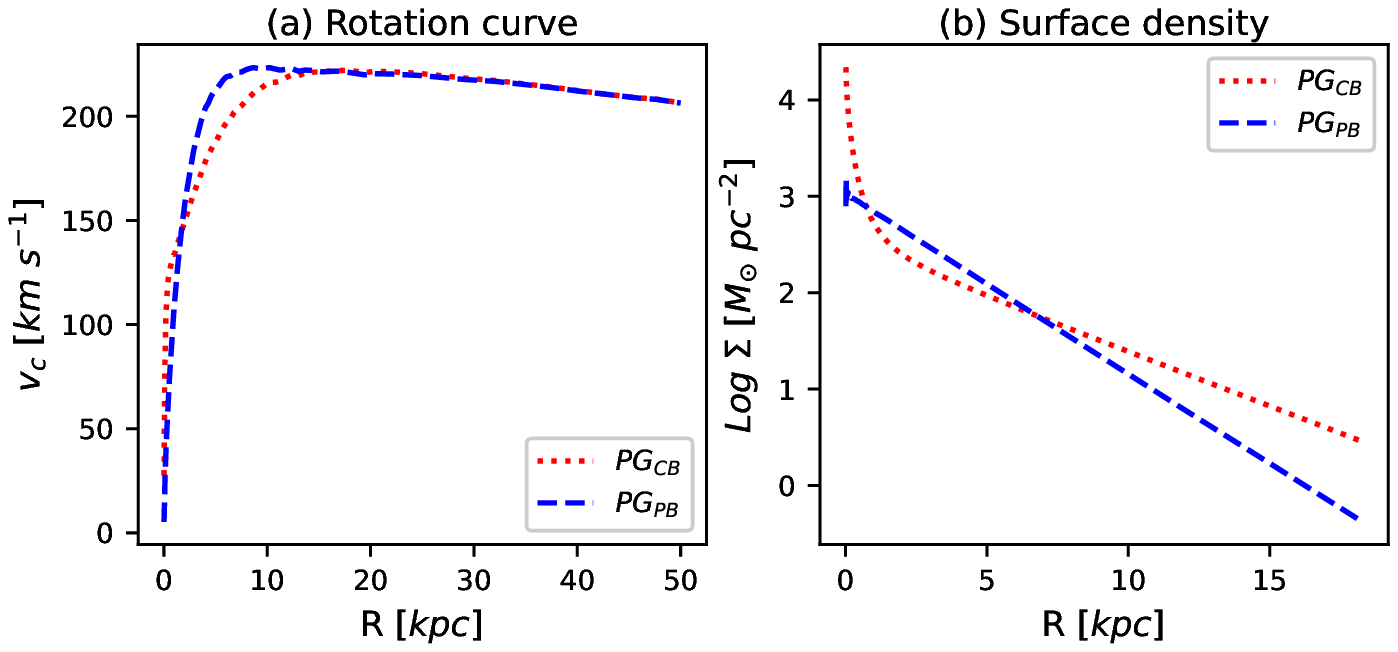}
    \caption{The left panel shows the total rotation curves of initial models of the major galaxy. The panel on the right shows the corresponding stellar surface densities of the galaxy. The red dotted curve and the blue dashed curve represent the $PG_{CB}$ and $PG_{PB}$ models respectively.}
    \label{fig:rot_curve_surf_den}
\end{figure}

To generate a bar stable major galaxy (hereafter referred to as an unbarred galaxy) and bar unstable major galaxy (hereafter referred to as a pseudobulge galaxy), we tuned the angular momentum fraction of the disk to halo, at a fixed disk scale height to scale radius ratio. This tuning of angular momentum decides the size of the disk (see \cite{Mo.Mao.White.1998} for the galaxy models). The unbarred galaxy models did not show any signatures of bar formation throughout the simulation time of 5~Gyrs but the barred galaxy forms the bar and buckles to form a pseudobulge. In fig.~\ref{fig:rot_curve_surf_den}, we have shown the rotation curves (left panel) and surface densities (right panel) of the major galaxy models.

The choice of the number of particles in the major galaxy is based on the region of interest within the galaxies. Since we are interested in the bulge and the disk of the major galaxy, we selected the number of particles in the major galaxy in such a way that the two body relaxation time is greater than the time of interest (i.e. the simulation time) in the central region of the galaxy \citep{Power2003}. The relaxation time is 5~Gyr for the models $PG_{CB}$ and $PG_{PB}$ at the radii 0.14~kpc and 0.27~kpc respectively. To confirm that our results do not depend on mass resolution of the simulations, we have also performed four high mass resolution simulation runs (one isolated and one $r_{p}=40~kpc$ for both $PG_{CB}$ and $PG_{CB}$ models) with twice as many particles as listed in the Table~1.

\begin{figure}
	\includegraphics[width=\columnwidth]{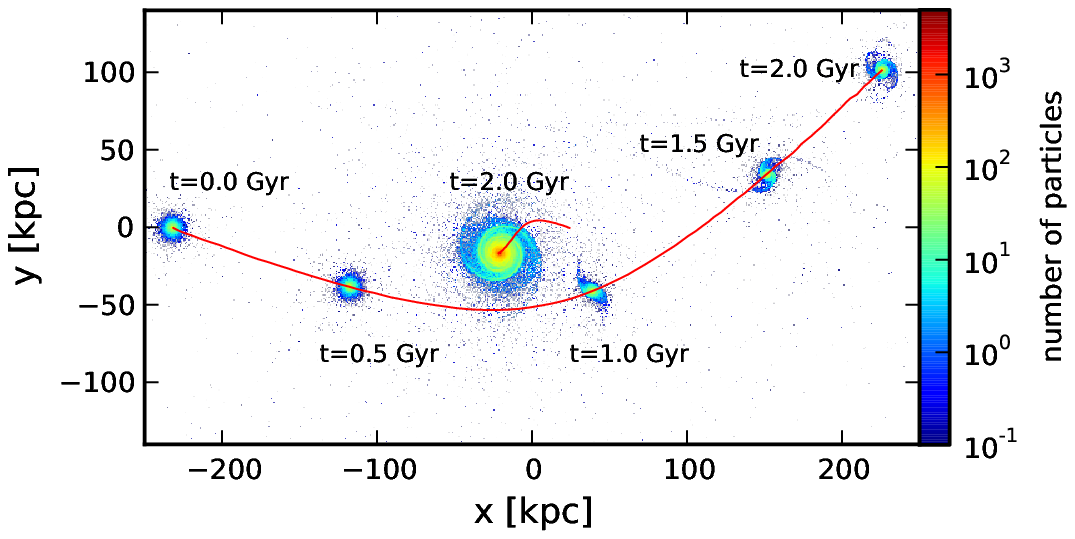}
    \caption{The orbits of two disk galaxies of mass ratio 10:1 undergoing a flyby interaction with pericenter distance of 40~kpc. The red solid curves trace the orbits of the galaxies during flyby. The minor galaxy is shown in time steps of 0.5~Gyr and the major galaxy is shown at the time 2~Gyr. The mass weighted particle number density is represented by the color on the log scale. In this illustration, only the stellar particles are shown.}
    \label{fig:orbits_of_galaxies}
\end{figure}

After generating the major and minor galaxies, we put them on hyperbolic orbits of eccentricity $e=1.1$ as illustrated in fig.~\ref{fig:orbits_of_galaxies} so that the minor galaxy does not decay by dynamical friction of the major galaxy. In this figure, the minor galaxy is shown at time intervals of 0.5~Gyr and the major galaxy is shown at time t=2.0~Gyr. The solid red curves traces the actual paths of the center of mass of the two galaxies in one of our simulations. The initial separation of the galaxies was chosen to be 255~kpc which is the sum of the virial radii ($r_{200}$) of the major and minor galaxies. We evolved the galaxies for 4~Gyr using the publicly available massively parallel code Gadget-2 \citep{Springel2001, Springal2005man, Springel2005}. The evolution time is set to 4.0~Gyr because galaxies are well separated after evolving for this much amount of time. In all of our simulations, the gravitational softening is 0.02~kpc for the stellar particles and 0.03~kpc for the halo particles. The change in total angular momentum is within 0.15$\%$ for all the simulations in 4~Gyrs of evolution, which ensures an output with minimum numerical errors. 

To study the effect of flyby interactions on the bulge and the disk of the major galaxy, we varied the distance of closest approach i.e. the pericenter distance of the galaxies ($r_{p}$) from 40~kpc to 80~kpc, keeping all the other parameters fixed. Hence, we have two sets of simulation models for each pericenter distance ($r_{p}$), which are : (1) a major galaxy with a classical bulge and (2) a major galaxy with a pseudobulge bulge. Each model is further simulated for 10:1 and 5:1 galaxy mass ratios. As a control model, we have also evolved the major galaxy in isolation for the same time period of 4~Gyr. The isolated model provides the benefit of removing the effect of secular evolution and the effect of discreteness, if any.


For naming the models, we have used pericenter distances determined assuming 2-body systems. For e.g. $r_{p}=40$ denotes the model with pericenter distance = 40~kpc as determined from a two body system. However, the real N-body systems always deviates from these pre-determined pericenter distances. Note that all the quantities are in units of the Hubble parameter '$h$' (where the Hubble constant is $H_{0}= 100$ $h$ $km$ $s^{-1} Mpc^{-1}$ and $h=0.67$ from \cite{Planck.Collaboration2020}) and all the plots are for the major galaxy unless otherwise specified.

\subsection{Analysis}
\label{sec:analysis}

We used the latest released version of GALFIT, which is a widely used two dimensional galaxy image decomposition tool, for the bulge-disk decomposition of the simulated galaxies \citep{Peng2003man, Peng2011}. Since we are interested in the effect of the flyby on the major galaxy or host galaxy only, we did the two dimensional image decomposition of the major galaxy and not the minor galaxy. Also, since GALFIT requires the galaxy images in `fits' format as an input, we first made `fits' format images of the galaxies using simulation snapshots at time steps of 0.2~Gyr. For the galaxy decomposition, we did not consider any background sky and hence the point spread function (psf) is a delta function. We used a simple exponential profile for the two dimensional fitting of a face-on disk,
\begin{equation}
    \Sigma_{d}(R)=\Sigma_{d0}\exp(-{\frac{R}{R_{s}}})
    \label{eqn:galfit_disk_face}
\end{equation}
where $\Sigma_{d0}$ is the central surface density of the disk and $R_{s}$ is the disk scale radius. For the two dimensional fitting of an edge-on disk, we used following profile
\begin{equation}
    \Sigma_{d}(R,z)=\Sigma_{d0}(\frac{R}{R_{s}})K_{1}(\frac{R}{R_{s}})\sech^{2}(\frac{z}{z_{0}})
    \label{eqn:galfit_disk_edge}
\end{equation}
where $K_{1}$ is a Bessel function and $z_{0}$ is the scale height of the disk. We used a sersic profile for the two dimensional fitting of the bulge
\begin{equation}
    \Sigma_{b}(R)=\Sigma_{b0}\exp[-b_{n}\{{(\frac{R}{R_{e}}})^{\frac{1}{n}}-1\}]
    \label{eqn:galfit_sersic}
\end{equation}
where $R_{e}$ is effective radius of the bulge, $\Sigma_{b0}$ is the surface density of the bulge at $R_{e}$, and $b_{n}$ is a function of the sersic index $n$. Previous studies of galaxy bulges suggest that the sersic index $n=2$ is a good proxy for distinguishing between classical bulges and pseudobulges \citep{Fisher2006, Fisher2008B}. Generally, classical bulges have a sersic index $n>2$ while pseudobulges have a sersic index $n<2$. We have applied this classification scheme for bulges in our study.

\begin{figure*}
	\includegraphics[width=0.85\textwidth]{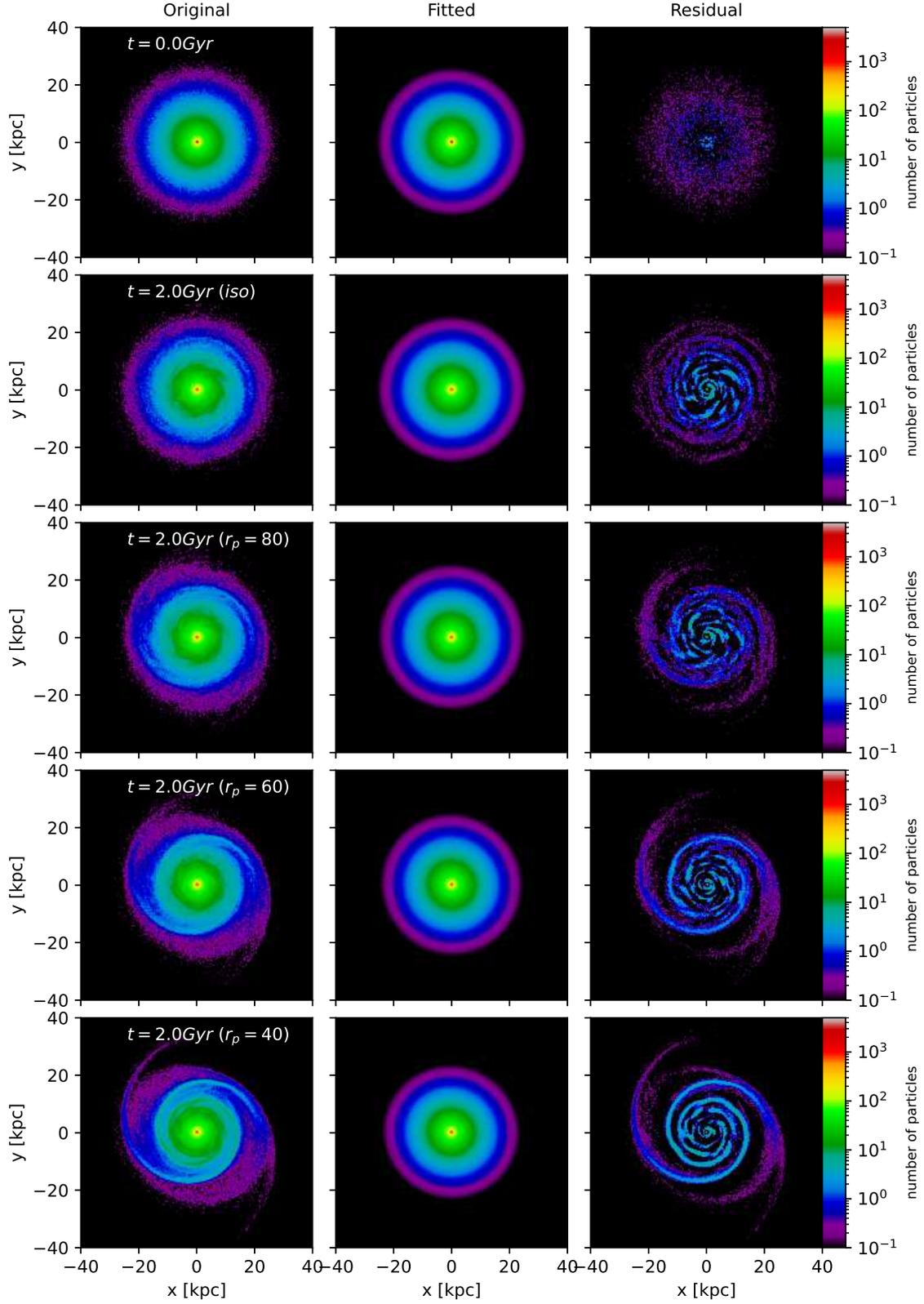}
    \caption{The two dimensional decomposition of the major galaxy with a classical bulge. The left, middle and, right columns show the original images, fitted images, and residual images respectively. The mass weighted particle number density is represented by the color on the log scale. For the sake of visualization, only positive pixels are shown in the residual images. From top to bottom, the five rows show the model galaxy at t=0.0~Gyr, $isolated$ model at t=2.0~Gyr, $r_{p}=80$ model at t=2.0~Gyr, $r_{p}=60$ model at t=2.0~Gyr, and $r_{p}=40$ model at t=2.0~Gyr respectively}
    \label{fig:2d_decomposition}
\end{figure*}

For improving the goodness of fit, we did several trials by setting different box sizes, bin sizes (pixel size), and convolution box sizes. We chose all these parameters in such a way that there is minimum effect of over-fitting and under-fitting. Therefore, we centered the major galaxy in a box of 80~kpc $\times$ 80~kpc size with a pixel size of 0.08~kpc $\times$ 0.08~kpc. For each fitting, the initial guess parameters of the galaxy components were chosen visually. See fig.~\ref{fig:2d_decomposition} for a visual illustration of the fitting procedure in face-on galaxies. The original, fitted and residual distributions of the simulated galaxies with classical bulges are shown in the first, second and third column respectively. Each row of the figure represents the two dimensional bulge-disk decomposition of a model galaxy at a given time as written in the legend of the first column. The figure suggests that GALFIT fits our simulated galaxies quite well with minimum residuals.

\begin{figure}
	\includegraphics[width=\columnwidth]{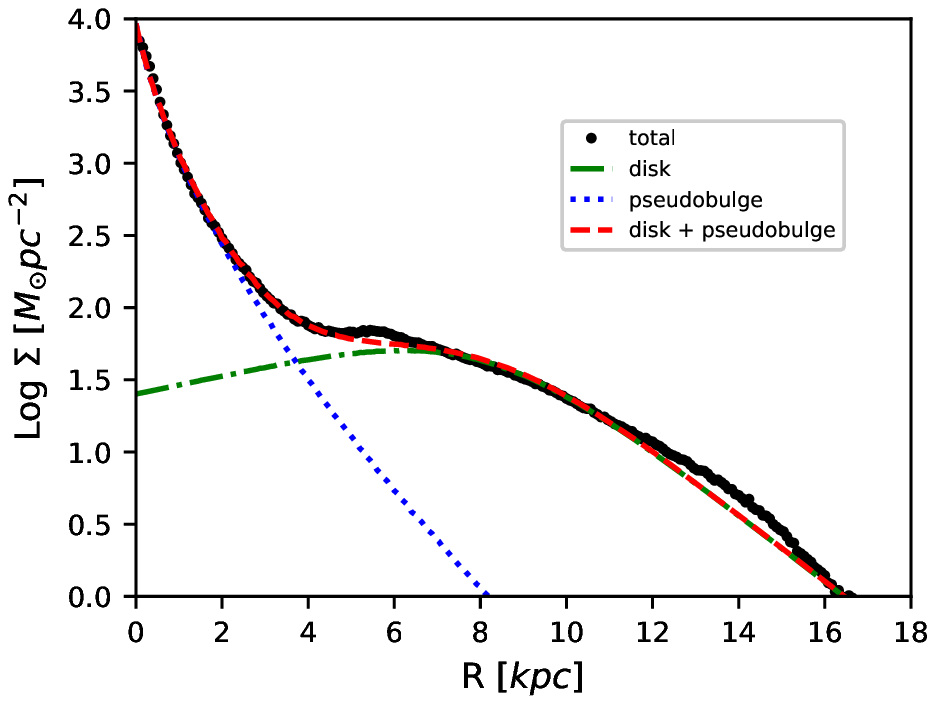}
    \caption{ One dimensional density profiles of pseudobulge galaxy model and fit obtained from its two dimensional decomposition. The black circle-solid curve, green dash-dotted curve, blue dotted curve, and red dashed curve demonstrate the total, disk, pseudobulge, and disk + pseudobulge profiles respectively. The disk profile is smoothly truncated in the inner region because most of the matter is swept by the bar. Our model pseudobulge fits well with $n=1.4$ sersic index.}
    \label{fig:decomp_profile}
\end{figure}

In our simulations, a pseudobulge is formed from the buckling of the bar. So the bar and pseudobulge are essentially the same component of the galaxy. Fig.~\ref{fig:decomp_profile} shows the 1-d density profiles of the pseudobulge galaxy model and the fits obtained from its 2d decomposition using GALFIT. It also includes the individual profiles of the disk and pseudobulge components. The disk profile is smoothly truncated in the inner region of the galaxy because most of the central mass is swept up by the bar. Our model pseudobulge fits well with $n=1.4$ sersic index which verifies its flat nature. However, the morphological decomposition of a galaxy reveals only the projected distribution of mass or light in the different components. This is fine for a spherical bulge but a flat bulge requires more detailed decomposition. Therefore, we explored kinematic methods to decompose the pseudobulges \citep{abadi.etal.2003}.

\begin{figure}
	\includegraphics[width=\columnwidth]{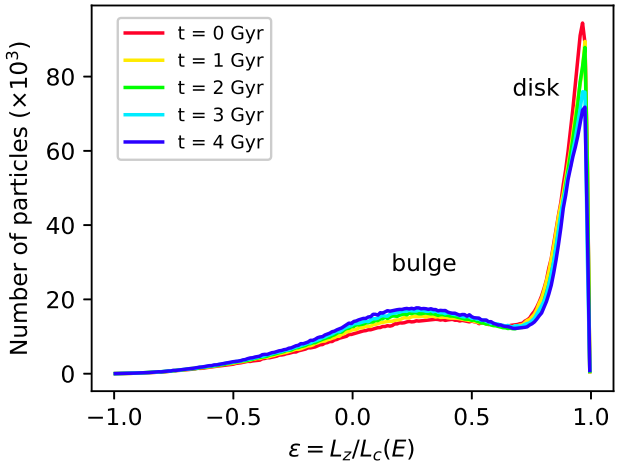}
    \caption{ An example of the kinematic decomposition of the bulge (hot) and disk (cold) components of the galaxy with a pseudobulge (model $PG_{PB}$). The X-axis denotes the ratio of stars' angular momentum perpendicular to the disk plane ($L_{z}$) and the maximum circular angular momentum ($L_{c}(E)$) corresponding to the stars' energy E. The peak around $\epsilon=1.0$ shows the rotation dominated component and the peak away from $\epsilon=1.0$ shows the dispersion dominated component. The different colors show the bulge-disk decomposition of the $isolated$ model with a pseudobulge ($PG_{PB}$) at time intervals of 1.0~Gyr.}
    \label{fig:kinematic_decomp}
\end{figure}

In the kinematic method, we used the angular momentum of the particles perpendicular to the galaxy plane to separate the `cold' disk particles and `hot' bulge particles that had larger velocity dispersion. Let $L_{z}$ be the angular momentum of a particle perpendicular to the galaxy plane, let $E$ be it's total energy, and $L_{c}(E)$ be the maximum angular momentum for the particle with energy $E$ in the galaxy potential. Now the ratio $|L_{z}/L_{c}(E)|$ will be close to one for cold/disk particles and close to zero for hot/bulge/halo particles (see the fig.~\ref{fig:kinematic_decomp}). For the calculation of $L_{c}(E)$, we divided all the particles in $\sim$200 equal size energy bins and found the particle with maximum $L_{z}$ in each bin. These $L_{z}$ are the $L_{c}(E)$ in the corresponding energy bins. After that, the $L_{c}(E)$ of all the particles are calculated using $3rd$ order spline interpolation of these $\sim$200 values.

All of our models show a two armed spiral structure during the flyby interactions, which are radially symmetric to each other (see the fig.~\ref{fig:2d_decomposition}). The spiral arms produced in flybys are very close to the logarithmic shape. So we used the Fourier analysis method as discussed in \cite{Sellwood1984} and \cite{Sellwood1986} to calculate the strength and pitch angle of the spiral arms. The Fourier coefficients are given by the following expression,
\begin{equation}
    A(m,p)=\frac{1}{N}\sum_{j=1}^{N} \exp[i(m\phi_{j}+p\ln R_{j})]
    \label{eqn:spiral_strenght}
\end{equation}
where $m$ is the number of spiral arms ($m=2$ for our models), $p$ is a real number related to the pitch angle ($\alpha$) of the spiral arms, $N$ is the number of stars in a given annular region from radii $R_{min}$ to $R_{max}$, and $(R_{j},\phi_{j})$ are the polar coordinates of the $j^{th}$ star. The range $R_{min}$ to $R_{max}$ is chosen in such a way that the maximum spiral arm strength lies in this range and no extra feature (e.g. bar and tidal arms) falls within this range. Then we calculated $A(m,p)$ for $p\in[-50,+50]$ at a step $dp=0.25$ and determine the parameter $p_{max}$ such that the value of $A(m,p)$ maximizes at $p=p_{max}$ \citep{Puerari.etal.2000}. So then the pitch angle is given by $\alpha=\arctan(m/p_{max})$ \citep{Sang-Hoon.etal.2015, Semczuk.etal.2017} and the spiral arm strength is given by $|A(m,p_{max})|$. For comparison, we also calculated the spiral arm strengths using the method as described for bars in equation~\ref{eqn:bar_strenght}. Both the methods give similar spiral arm strengths.

\section{Results}
\label{sec:results}
We have simulated flyby interactions for galaxies with mass ratio 10:1 and 5:1, and for two types of bulges in the major galaxy: the classical bulge and the pseudobulge. The pericenter distances are set at 40~kpc, 60~kpc, and 80~kpc. As a control case, we have evolved isolated galaxy models for the same length of time. In the following subsections we discuss our results of the effect of galaxy flybys on the disks and bulges. As mentioned earlier, we focus on the properties of the major galaxy only, and not the minor one.

\subsection{Disk Scale Radius}
\label{sec:disk_radius}

\begin{figure}
	\includegraphics[width=\columnwidth]{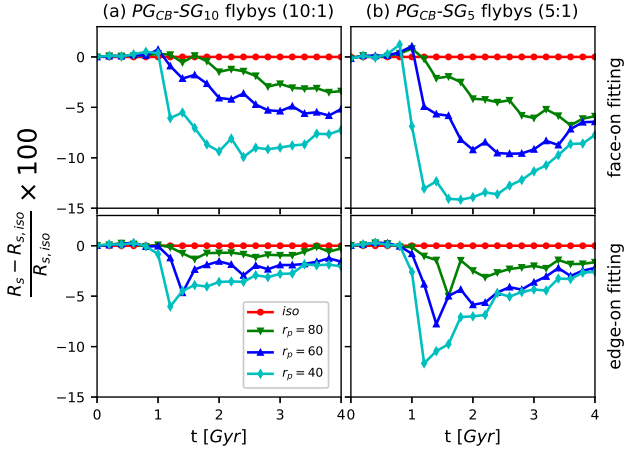}
    \caption{The evolution of the percentage change in the disk scale radius of the unbarred, classical bulge galaxy due to flybys. The left and right columns show the flyby simulations of $1/10$ and $1/5$ galaxy mass ratios respectively. The top and bottom rows represent the outputs from face-on and edge-on galaxy fitting respectively. The red circle-solid curves, the green down triangle-solid curves, the blue upper triangle-solid curves, and the cyan diamond-solid curves represent the models $isolated$, $r_{p}=80~kpc$, $r_{p}=60~kpc$, and $r_{p}=40~kpc$ respectively.}
    \label{fig:disk_radius}
\end{figure}

During close flybys, galaxies experience strong tidal forces due to each others gravity. This tidal force pulls out the stellar mass from the galaxies and results in the formation of strong spiral arms and tidal streams as shown in the third column of fig~\ref{fig:2d_decomposition}. Since the spiral arms form from the stellar particles of the pristine disk, there will be some resultant changes in the disk surface densities. The question is how does the formation of spiral arms affect the pristine disk? To answer this question, we examined the time evolution of the percentage change in the disk scale radius ($R_{s}$) of the unbarred, classical bulge galaxies in flybys with 10:1 and 5:1 mass ratios, relative to a control isolated model, as shown in column (a) and column (b) of fig~\ref{fig:disk_radius}. The top and bottom rows are the corresponding face-on and edge-on fits of the galaxies. We have performed the edge-on fitting of the galaxies for two perpendicular viewing angles to minimize the bias of the viewing angle on the morphology of the galaxy. The bottom row shows the percentage change in the average disk scale radius for the edge-on fitting. At the beginning of the simulation, the face-on and edge-on fittings give the disk scale radius to be 3.65~kpc and 3.78~kpc respectively. The scale radius from edge-on fitting is nearly equal to the theoretical scale radius (3.8~kpc) but the face-on fitting gives slightly smaller disk scale radius. The cause of this mismatch is the discreteness of the matter distribution in the simulations, which lacks a smooth distribution of particles in the outermost region. Therefore the outermost particles are excluded in the face-on fitting but edge-on fitting does not have this drawback. However, this small mismatch is not a problem for studying the disk evolution in our simulations.

From this figure (fig~\ref{fig:disk_radius}) it is very clear that the scale radius of the unbarred, classical bulge galaxy decreases with time. The magnitude of the decrease in disk scale radius depends on the pericenter distances of the galaxies. The smaller the pericenter distance, the larger is the change in disk scale radius. Note that the isolated models do not show any change in the scale radius because they are control models and all the change are measured with respect to them. The closest flyby models, which experience the largest tidal forces, show the greatest change in the scale radius because they also have the strongest spiral features. Hence the changes in the disk scale radii are correlated with the formation of the spiral arms (which are discussed later in this section). All the models show a significant change in the disk scale radius just after the pericenter passage, at approximately t=1.0~Gyr. Before pericenter passage, all the models show no significant change in the disk scale radius. In both 10:1 and 5:1 flybys, the face-on fittings show a larger change compared to the edge-on fitting. This is due to the presence of spiral arms which remain in the residual in the face-on fits. However, in the edge-on fits a significant fraction remains in the disk. For a given pericenter distance ($r_{p}$), the 5:1 simulations show more changes than the 10:1 simulations. Similarly, for a given major to minor galaxy mass ratio, the closest pericenter passage shows more change than the relatively farther pericenter passage. The effects of close flybys are short lived. As the minor galaxy goes away, the matter pulled out due to the flyby falls back into the major galaxy and starts building the disk again.

\subsection{Disk Scale Height or Thickness}
\label{sec:disk_height}

\begin{figure}
	\includegraphics[width=\columnwidth]{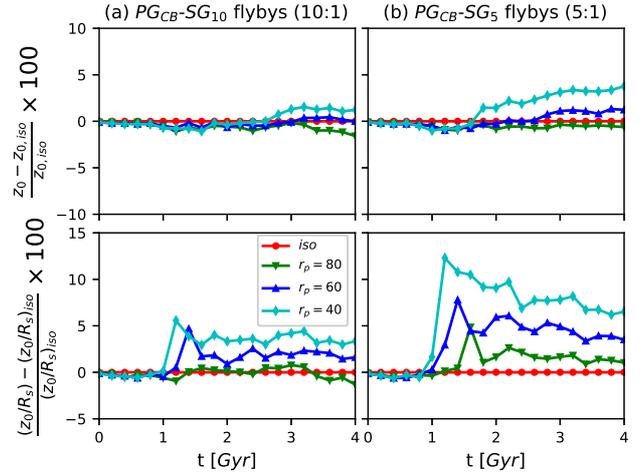}
    \caption{Evolution of the percentage change in the disk height or thickening in unbarred galaxies with bulges due to flybys. Left and right columns show the flyby simulations of galaxies with $1:10$ and $1:5$ mass ratio respectively around the major galaxy with classical bulge. Top and bottom rows represent the change in the disk scale height and the ratio of the disk scale height to the disk scale radius respectively. The red circle solid curves, green down triangle solid curves, blue upper triangle solid curves, and cyan diamond solid curves represent the models $isolated$, $r_{p}=80~kpc$, $r_{p}=60~kpc$, and $r_{p}=40~kpc$ respectively.}
    \label{fig:disk_height}
\end{figure}

The minor flyby interactions of galaxies heats up the minor galaxy dynamically and can also induce disk instabilities (e.g. bar formation and disk warping) in the minor galaxy which results in the thickening of the disk \citep{lokas.etal.2014, Gajda.etal.2018, Lokas.etal.2018}. Although the minor galaxy exerts a much weaker tidal force on the major galaxy, the perturbation may dynamically heat up the major galaxy disk as well. To see if there is any disk thickening for our simulations of flybys with unbarred, classical bulge galaxies, we have plotted the percentage change in the disk scale height ($z_{0}$) and the ratio of disk scale height to disk scale radius ($\frac{z_{0}}{R_{s}}$) relative to the control isolated model, as a function of time in the top and bottom rows of the fig.~\ref{fig:disk_height} respectively. The left (a) and right (b) columns of the figure represent the galaxies in 10:1 and 5:1 mass ratio flybys respectively. To reduce the bias of the viewing angle on the morphology of the galaxies, we have plotted average disk scale heights and the ratio of the average disk scale height to average scale radius of two fitted models at perpendicular viewing angles. From the first row of the figure, we can see that both the models show very small or insignificant disk thickening due to the dynamical heating by the minor flybys. The model $r_{p}=40~kpc$ for a mass ratio of 5:1 shows a maximum of $\sim 4\%$ increase in the disk scale height due to the flyby. These results indicate that minor flybys cannot heat or thicken the disks of the major galaxy significantly. Only the very close and nearly equal mass flybys (major flyby) can heat the disk of the major galaxy significantly.

However, the disk thickness can also be measured relative to the disk scale radius at the time of measurement as shown in the bottom row of the figure (fig.~\ref{fig:disk_height}). To bring out the effect of flybys on the disk thickening, we have shown the percentage change in the ratio of the disk scale height to the disk scale radius relative to that of control isolated model. At the beginning of the simulation, the disk scale height is $0.1R_{s}$ which is equal to the theoretical value we started with. As we go forward in time, all the models irrespective of the flyby mass ratio follow the same track until nearly t=1.0~Gyr i.e. before close approach. After that, all the models move on different tracks. In the 10:1 models, the disk thickening remains nearly the same after the passage of the minor galaxy, but the 5:1 models show sudden disk thickening just after the flyby, and then thinning due to increasing disk scale length as discussed in subsection~\ref{sec:disk_radius}. From both panels of the bottom row, it is clear that the closest approach produces the largest effect on disk thickening at a given major to minor galaxy mass ratio. Similarly, the major flyby produces the greatest effect on disk thickening at a given pericenter passage.

\subsection{Spiral Arms Strength}
\label{sec:spiral_arms}

\begin{figure}
	\includegraphics[width=\columnwidth]{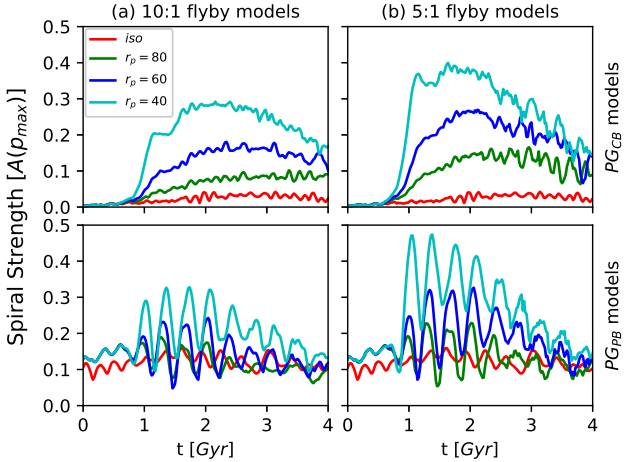}
    \caption{Time evolution of the spiral arms strength. The left and right columns show the flyby simulations of mass ratios 10:1 and 5:1 respectively. The top and bottom rows represent the major galaxy model of classical bulge and pseudobulge respectively. The red, green, blue, and cyan curves represent the models $isolated$, $r_{p}=40~kpc$, $r_{p}=60~kpc$, and $r_{p}=80~kpc$ respectively.}
    \label{fig:spiral_arms}
\end{figure}

The tidal force of the minor galaxy pulls out stellar particles from the major galaxy. At the same time, the resonance between the stellar orbits and the orbit of the minor galaxy helps the major galaxy to develop spiral arms. In our {simulations, the} galaxies move in a prograde-prograde configuration which is the most favourable orientation for the formation of spiral arms. All of our models develop two spiral arms during the interactions which are radially symmetric to each other (see the fig.~\ref{fig:2d_decomposition}). The strength of spiral arms in a galaxy represents how well the spiral features are distinguishable from the host disk. Hence the amplitude of the m=2 Fourier mode will be a good estimator of the spiral arm strength at any given radius (see equation~\ref{eqn:spiral_strenght}).

In fig~\ref{fig:spiral_arms}, we have plotted the strength of the spiral arms for our models. The left and right columns of the figure show the strength of spiral arms in the flyby models with mass ratios 10:1 and 5:1 respectively. The top and bottom rows represent the spiral arms strength in galaxies with a classical bulge and a pseudobulge respectively. From all the panels of the figure, one can see that the major galaxy in all the flybys shows a sudden increase in the spiral strength at approximately t=1.0~Gyr which is the time of pericenter passage of the galaxies. There is no change or negligible change in the strength of the spiral arms for the control or isolated models throughout the simulation. This shows the importance of flyby interactions in the formation of strong spiral arms.

The time evolution of the spiral arms strength in the classical bulge models shows that the strength increases as the satellite crosses the pericenter, which is at approximately 1.0~Gyr. After some time, the magnitude of the arm strength reaches its maximum and starts decreasing slowly. As expected the value of the peak strength increases with decreasing pericenter distance of the galaxies. Also, at a given pericenter distance, the 5:1 flyby models show higher peak strength than the 10:1 flyby models. The closer flybys attain peak strength earlier than the farther flybys. However, although the closest flyby interactions show the most rapid growth in spiral arms strength, they also show rapid decay in the spiral arms strength. As a result all the flyby models reach the same strength level after some time. This result shows the importance of satellite galaxies or massive clusters as spiral strength boosters. They help the major galaxies to maintain well defined and long lasting spiral arms.

Also, for the close flybys of the unbarred, classical bulge models, the spiral arms can be traced all the way to the center of the galaxy. For example, in the fig.~\ref{fig:2d_decomposition}, the $r_{p}=40~kpc$ model shows a grand design spiral structure after the flyby. The two spiral arms are distinct all the way to the center. It is well known that the strength of global disk instabilities in disks (e.g. bars, spiral arms) depend on the mass and the concentration of the bulge \citep{shen.etal.2004, Athanassoula.Lambert2005, kataria.das.2018}. We see this effect in our simulations as well, since both the strength and the extent of the spiral arms formed in the flybys depends on the distance of closest approach and the presence of the bulge mass.

The evolution of spiral arms strength in the pseudobulge models is even more interesting as can be seen from the bottom row of the fig~\ref{fig:spiral_arms}. Both the 10:1 and 5:1 flyby models exhibit an oscillatory nature on top of the time-varying spiral arms strength, whose nature is similar to those for classical the bulge models. After the pericenter passage, the crest values of these oscillations vary in an approximately similar manner as the spiral arm strength of the classical bulge models. They first increase, reach a peak value and then start decreasing slowly. At the end of the interaction, at 4~Gyr, all the models show similar spiral arm strengths which are equal to the control or isolated model. The isolated models have some initial arms strength $A(p_{max})=0.1$ which is because we have taken the initial pseudobulge models after the evolution of bar strength becomes nearly time independent (see the fig.~\ref{fig:bar_strength}), and by then these models have grown weak spiral arms. All the flyby models show the same oscillation frequency, irrespective of the flyby mass ratio and pericenter distances. This indicates that these oscillations are intrinsic to the host galaxy and the flyby interactions amplify the amplitudes of these oscillations. Barred galaxies have intrinsic resonances \citep{Binney.Tremaine.2008}. The oscillations in the effective potential and the positions of equilibrium points near the co-rotation resonances in barred galaxies has been studied by \cite{Wu.etal.2016}.

\subsection{Sersic Index of the Classical Bulge}
\label{sec:sersic_index}

\begin{figure}
	\includegraphics[width=\columnwidth]{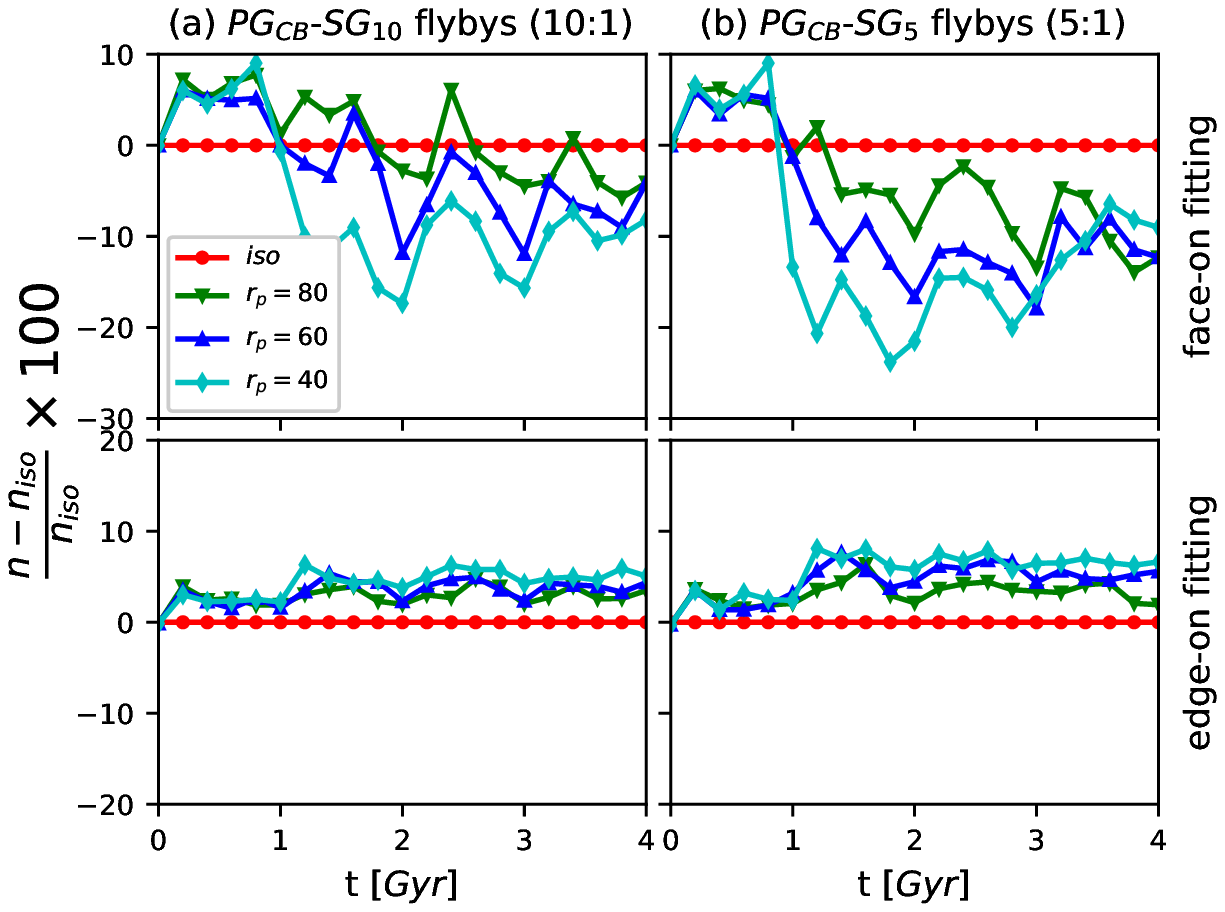}
    \caption{ Evolution of the percentage change in the sersic index of the bulge in unbarred galaxies due to flybys. Left and right columns show the flyby simulations of $1/10$ and $1/5$ mass minor galaxy respectively around the major galaxy with classical bulge. Top and bottom rows represent the outputs from face-on and edge-on galaxy fitting respectively. The red circle solid curves, green down triangle solid curves, blue upper triangle solid curves, and cyan diamond solid curves represent the models $isolated$, $r_{p}=80~kpc$, $r_{p}=60~kpc$, and $r_{p}=40~kpc$ respectively.}
    \label{fig:sersic_index}
\end{figure}

The evolution of the sersic indices of the classical bulges relative to the control isolated model, derived from the two dimensional decomposition of the major galaxy, is shown in fig.~\ref{fig:sersic_index}. As in the previous figures, the left and right columns represent flybys of 1:10 and 1:5 mass ratios. The output from the face-on and edge-on bulge-disk fittings are shown in the top and bottom rows respectively. Our classical bulge models have initial sersic index values ($n$) greater than 2. It should be noted that the value of $n$ strongly depends on the surface density of the host galaxy disk i.e. the background of the bulge. For example, the classical bulge without a disk has a sersic index of $n=3.6$ but when the disk is included the face-on fit gives the value $n=2.1$. However, with the edge-on disk, the sersic index remains approximately equal to that without a disk. In our study, we fitted both components of the galaxy simultaneously because that is how it is done in observational studies.

\begin{figure}
	\includegraphics[width=\columnwidth]{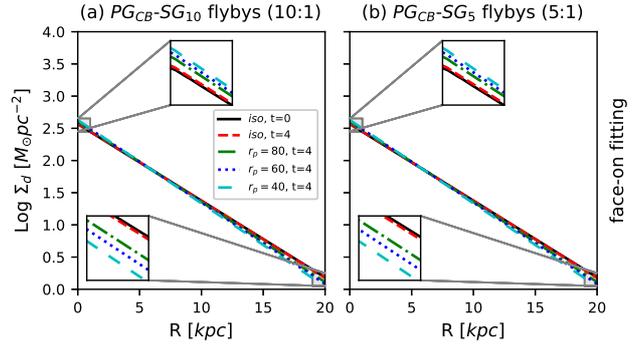}
    \caption{The left and right panels show the initial and final disk surface densities from face-on decomposition of the unbarred galaxies of 10:1 and 5:1 flyby models respectively. The black solid curves show the initial profiles and red dashed curves, green dash-dotted curves, blue dotted curves, and cyan dash-spaced curves represent the final profiles of the $isolated$, $r_{p}=80~kpc$, $r_{p}=60~kpc$, $r_{p}=40~kpc$ models respectively. The rectangular insets show the 5 times zoom in regions.}
    \label{fig:disks_density}
\end{figure}

In the face-on fittings, the value of the sersic index increases until $t\sim$1.0~Gyr for all the models (top row of the fig.~\ref{fig:sersic_index}). This small increment in the sersic index is due to the steepening of the surface density of the inner disk region \citep{Hohl1971, Laurikainen2001, Debattista2006} which contributes to the mass of the spherical/classical bulge. After t=1.0~Gyr, the value of the sersic index decreases for all the models. The decrement of the sersic index after 1.0~Gyr, which is the time at when the minor galaxy passes the pericenter, is due to the formation of strong spiral arms. The spiral arms reduce the surface density of the outer disk region. For the sake of visualisation, we have plotted the final and initial disk surface densities from the face-on decomposition of unbarred, classical bulge galaxies of 10:1 and 5:1 flybys models in the left and right panels of fig.~\ref{fig:disks_density} respectively. The density change in the inner and outer disk regions causes the bulge-disk fitting to take some bulge particles into the disk component, thus reducing the value of the sersic index. In both flyby models, the change in the sersic index just after pericenter passage is the largest for the closest flyby. But it become nearly equal in the end for all the models of the given galaxy mass ratio. In the end, the 5:1 simulations show more change than the 10:1 simulations. Hence it appears that closer and major flybys are more effective in transforming classical bulges into pseudobulges compared to more distant and minor flybys.

\begin{figure}
	\includegraphics[width=\columnwidth]{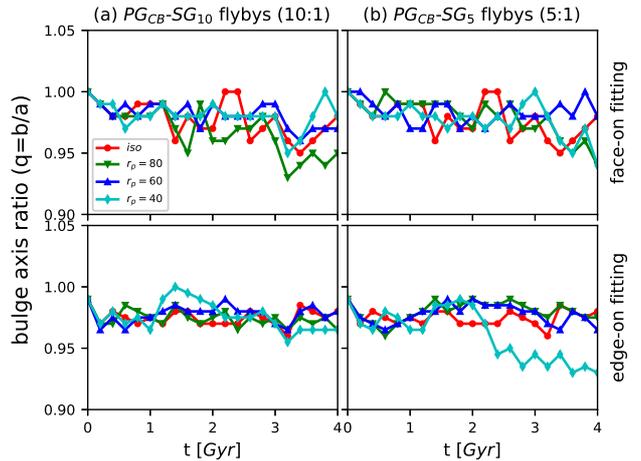}
    \caption{Evolution of the minor to major axis ratio ($q=b/a$) of the classical bulges in unbarred galaxies. Left and right columns show the 10:1 and 5:1 flybys simulations respectively. Top and bottom rows represent the outputs from face-on and edge-on galaxy fitting respectively. The red circle solid curves, green down triangle solid curves, blue upper triangle solid curves, and cyan diamond solid curves represent the models $isolated$, $r_{p}=80~kpc$, $r_{p}=60~kpc$, and $r_{p}=40~kpc$ respectively.}
    \label{fig:bulge_axis_ratio}
\end{figure}

However, this apparent decrement in the sersic index due to flyby interactions does not really represent the flattening of the classical bulge. This is because if the bulge became more disky or flattened in the z direction, it would be effectively turning into a pseudobulge or it would have a more oval shape. However, we could not detect any such change in the bulge shape in any of the classical bulge models. Fig.~\ref{fig:bulge_axis_ratio} illustrates this result. It shows the time evolution of the minor to major axis ratio ($q=b/a$) of the classical bulges. The left and right columns represent the 10:1 and 5:1 flyby models respectively. The top and bottom rows show the outputs of face-on and edge-on fitting respectively. All the panels show nearly constant axis ratios for the bulges. But the two dimensional decomposition of the face-on galaxy gives a lower sersic index which corresponds to the flattening of the bulge. This lowering of the sersic index, however, is due to the changing mass distribution in the disk caused by the flyby interaction as discussed earlier in this subsection. Hence, we conclude that classical bulge shapes are not changing due to flyby interactions.

In the edge-on fittings (see the bottom row of the fig.~\ref{fig:sersic_index}), in contrast to the face-on fittings, the value of the sersic index for all the flyby models is always greater than or equal to the corresponding isolated model. Hence, the flybys are not decreasing the sersic index of the classical bulge (or flattening the spherical bulge). Therefore, the real indicator of bulge type/shape is the edge-on decomposition of the galaxy.

\subsection{Mass of the Classical Bulge}
\label{sec:bulge_mass}

\begin{figure}
	\includegraphics[width=\columnwidth]{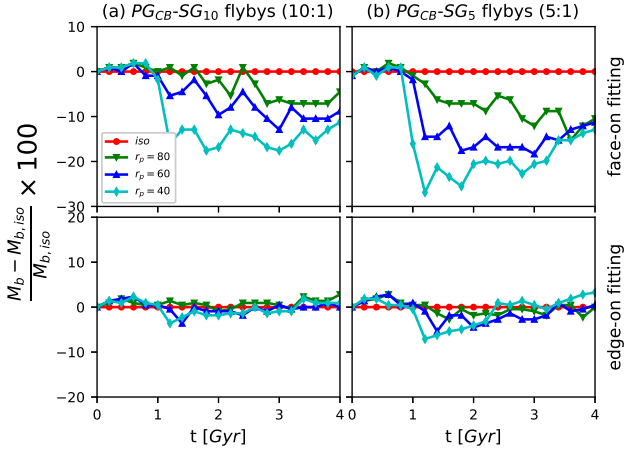}
    \caption{Evolution of the percentage change in the mass of the bulge in unbarred galaxies due to flybys. Left and right columns show the flyby simulations of $1/10$ and $1/5$ mass minor galaxy respectively around the major galaxy with classical bulge. Top and bottom rows represent the outputs from face-on and edge-on galaxy fitting respectively. The red circle solid curves, green down triangle solid curves, blue upper triangle solid curves, and cyan diamond solid curves represent the models $isolated$, $r_{p}=80~kpc$, $r_{p}=60~kpc$, and $r_{p}=40~kpc$ respectively.}
    \label{fig:bulge_mass}
\end{figure}

To see whether flybys affect the mass of classical bulges or not, we calculated the bulge mass at time steps of 0.2~Gyr using the integrated magnitudes of the bulges obtained from the output file of GALFIT. The left and right columns of fig.~\ref{fig:bulge_mass}, show the time evolution of the percentage change in the bulge mass of unbarred, classical bulge galaxies for the 10:1 and 5:1 flyby models respectively. The output from face-on and edge-on fittings are shown in the top and bottom rows of the figure respectively. The initial mass of the bulge in the face-on fitting (0.42 $\times 10^{10}M_{\odot}$) is smaller than the edge-on fitting (0.55 $\times 10^{10}M_{\odot}$). The bulge mass from edge-on fitting is much closer to the initial value of 0.6 $\times 10^{10}M_{\odot}$. In the face-on fittings, there are clear evolutionary trends which are similar to the evolutionary trends of the sersic indices. There is a marginal increase in the bulge masses until the time t=1.0~Gyr of the evolution. This marginal increment is due to the steepening of the surface density of the inner disk region \citep{Hohl1971, Laurikainen2001, Debattista2006} which contributes to the mass of the bulge. Then there is some decrease in the bulge masses for both types of flyby models after 1.0~Gyr of the evolution. This decrease in the bulge mass is due to the formation of strong spiral arms after the flyby interaction. The formation of spiral arms reduces the surface density in the outer region of the disk (see fig.~\ref{fig:disks_density}). Hence, the redistribution of disk particles causes some of the bulge particles to fit into the disk which changes the mass of the bulge in the 2d bulge-disk decomposition of the face-on galaxies. For a given pericenter distance, the 5:1 flybys show more change than the 10:1 flybys. Similarly, the closest pericenter passage shows the more change at a given major to minor galaxy mass ratio. But at the end of the simulations of a given galaxy mass ratio, the change in all the models appear to converge to the same value. On the other hand, edge-on fittings do not show any change in the bulge mass for both the flyby models as seen in the bottom row of the figure. Thus we conclude that flybys do not effect the mass of the classical bulges in the major galaxy.

\subsection{Angular Momentum of the Classical Bulge}
\label{sec:classical_bulge_Lz}

\begin{figure}
	\includegraphics[width=\columnwidth]{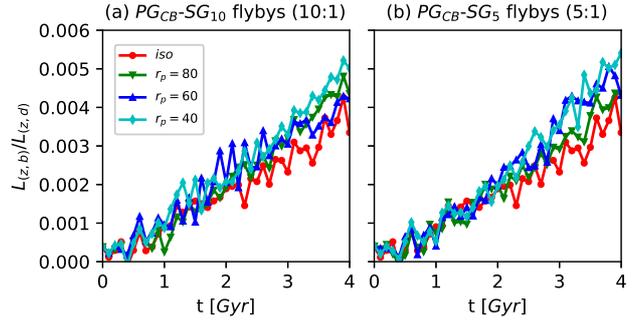}
    \caption{Evolution of the bulge to disk angular momentum ratio within a sphere of 5 bulge effective radius ($5R_{e}$) in unbarred galaxies. Left and right panels show the 10:1 and 5:1 flyby models respectively. The red circle solid curves, green down triangle solid curves, blue upper triangle solid curves, and cyan diamond solid curves represent the models $isolated$, $r_{p}=80~kpc$, $r_{p}=60~kpc$, and $r_{p}=40~kpc$ respectively.}
    \label{fig:classical_bulge_Lz}
\end{figure}

Flybys play an important role in the angular momentum transfer between galaxies. They can spin-up or spin-down the galaxies depending on the configuration of the flyby, such as prograde-prograde, prograde-retrograde, retrograde-prograde, or retrograde-retrograde directions of the galaxies orbits \citep{Bett.Frenk.2012}. But how do the bulges of the galaxies respond to the angular momentum transfer in flybys? To answer this question, we have calculated the z-component of the angular momentum of the bulge and disk within a sphere of radius 5 times the bulge effective radius ($5R_{e}$). In fig.~\ref{fig:classical_bulge_Lz}, we have shown the time evolution of the bulge to disk angular momentum ratio in unbarred, classical bulge galaxies. The left and right panels of this figure show the 10:1 and 5:1 flyby models respectively. We have plotted only the ratio of the z-component of the angular momenta because all of our model galaxies lie in the x-y plane.

From the two panels of the fig.~\ref{fig:classical_bulge_Lz}, we can clearly see that the ratio of bulge to disk angular momentum is increasing with time in all the models. This is the indication of the angular momentum transfer from the disk to bulge. However, until 2~Gyr, all the flyby models show a gain similar to that of the isolated model. There is a small angular momentum gain in the flyby models after 2~Gyr but it is not significant enough to clearly distinguish an angular momentum gain in the flyby models compared to the corresponding isolated model. Although it is possible that multiple prograde flybys may add up to a significant transfer of angular momentum to the bulge, it must however be noted that in nature, flybys are not always on prograde orbits. Hence, classical bulges do not gain significant angular momentum in flybys, but instead they could gain angular momentum due to the rotating disk \citep{Pedrosa.Tissera.2015}. The classical bulges in galaxies reside in the deepest part of the potential and there in these models there is no bar type instability in their disks. Hence they do not gain much angular momentum. However, the case is very different when the galaxies are barred as there can be significant angular momentum gain by the bulge \citep{kataria.das.2018}.

\subsection{Evolution of the pseudobulge}
\label{sec:pb_evolution}

\begin{figure}
	\includegraphics[width=\columnwidth]{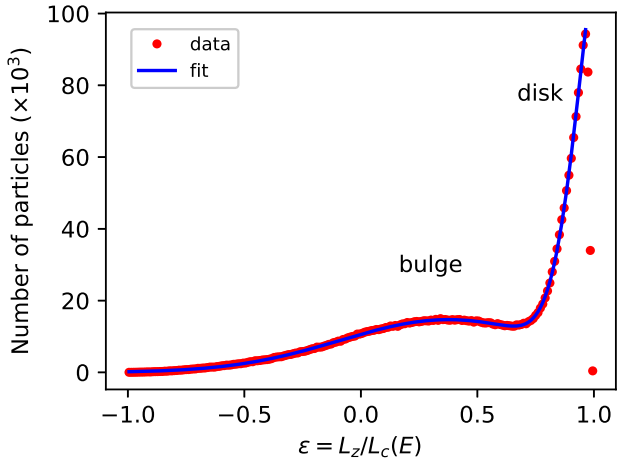}
    \caption{ An example of two generalized Gaussian fitting (up to the disk peak) to the kinematic decomposition of bulge and disk components of the galaxy with pseudobulge ($PG_{PB}$). The X-axis shows the ratio of stellar particle angular momentum perpendicular to the disk plane ($L_{z}$) and the maximum circular angular momentum ($L_{c}(E)$) corresponding to stellar energy E. The peak around $\epsilon=1.0$ shows the rotation dominated component and the peak away from $\epsilon=1.0$ shows the dispersion dominated component. The red filled circles and blue continuous curve show the data points and best fit curve respectively.}
    \label{fig:kinematic_decomp_fit}
\end{figure}

As discussed before in the subsection~\ref{sec:analysis}, the existence of the bar and the boxy/peanut pseudobulge makes the 2d decomposition of the galaxy very difficult. Therefore, we have used the kinematic decomposition method for the models with pseudobulges ($PG_{PB}$) \citep{abadi.etal.2003}. For the kinematic decomposition of the bulge, stellar particles within 15~kpc from the center of the galaxy are used because the bulge resides in the center of the galaxy. We simultaneously fitted the peaks corresponding to the bulge and the disk in the $\epsilon=L_{z}/L_{c}(E)$ distribution plot using two generalized Gaussians (for example see the fig.~\ref{fig:kinematic_decomp_fit}). We used the position ($\epsilon_{mean}$) and full width at half maximum ($\epsilon_{fwhm}$) of the generalized Gaussian corresponding to the bulge component as the indicator of bulge evolution. The idea is that the $\epsilon_{mean}$ of the bulge will move toward $\epsilon=0$ on increasing the random motion of the stellar particles, which represents the dynamical heating of the bulge, and vice-versa for dynamical cooling which is due to increasing ordered motion of the stellar particles. Similarly, the $\epsilon_{fwhm}$ will be larger for bulges that are less distinguishable from the disk compared to small $\epsilon_{fwhm}$ values which represent bulges that are dominant in disks.

\begin{figure}
	\includegraphics[width=\columnwidth]{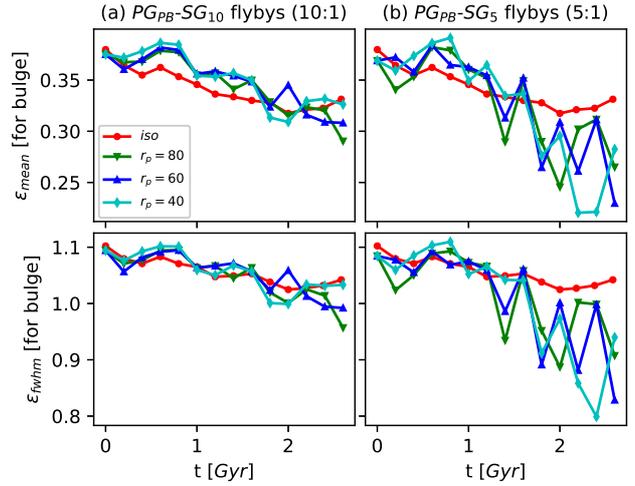}
    \caption{The top and bottom rows show the time evolution of the position ($\epsilon_{mean}$) and full width at half maximum ($\epsilon_{fwhm}$) of the generalized Gaussian fitted to the pseudobulge component. Left and right columns represent the flyby simulations of $1/10$ and $1/5$ mass minor galaxy respectively around the major galaxy with pseudobulge. The red circle solid curves, green down triangle solid curves, blue upper triangle solid curves, and cyan diamond solid curves represent the models $isolated$, $r_{p}=80~kpc$, $r_{p}=60~kpc$, and $r_{p}=40~kpc$ respectively.}
    \label{fig:pb_mean_fwhm}
\end{figure}

The top and bottom rows of fig.~\ref{fig:pb_mean_fwhm} demonstrate the time evolution of the mean position ($\epsilon_{mean}$) and full width at half maximum ($\epsilon_{fwhm}$) of the bulge component in galaxies with pseudobulges ($PG_{PB}$) respectively. The left and right columns show the outputs from 10:1 and 5:1 flyby models respectively. From the left column of this figure, one can easily interpret that there is no effect of the 10:1 flyby interactions on the pseudobulge of the major galaxy. The time evolution of $\epsilon_{mean}$ and $\epsilon_{fwhm}$ in the 10:1 flyby models are approximately similar to that of the corresponding isolated models. Although these flyby models do show some increase in the $\epsilon_{mean}$ during the closest pericenter passage, it does not last long. Hence, we can conclude that flybys with relatively low mass satellites cannot affect the evolution of the pseudobulge of the host galaxy.

In contrast to the 10:1 flyby simulations, the 5:1 flyby simulations show significant deviation from the corresponding control isolated models, as can be seen from the right column of the fig.~\ref{fig:pb_mean_fwhm}. The values of $\epsilon_{mean}$ and $\epsilon_{fwhm}$ decrease relative to the corresponding isolated models after the passage of the minor galaxy. The decrease in  $\epsilon_{mean}$ indicates increasing random motion in the bulge i.e the bulge becomes kinematically hotter. On the other hand, decrease in the $\epsilon_{fwhm}$ value indicates that the bulge is becoming more distinct from the disk. These results show that flybys with relatively large mass ratios (such as 1:5) can increase the random motion of stars in the pseudobulge of the host galaxy.

\begin{figure}
	\includegraphics[width=\columnwidth]{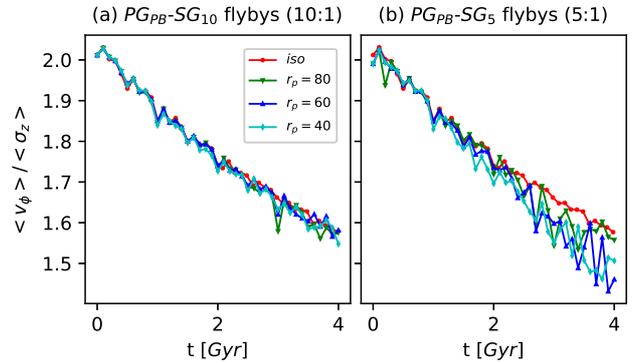}
    \caption{ The time evolution of the ratio of the mean azimuth velocity to mean vertical dispersion ($<v_{\phi}>/<\sigma_{z}>$) within 15~kpc. Left and right columns represent the flyby simulations of $1/10$ and $1/5$ mass minor galaxy respectively around the major galaxy with pseudobulge. The red circle solid curves, green down triangle solid curves, blue upper triangle solid curves, and cyan diamond solid curves represent the models $isolated$, $r_{p}=80~kpc$, $r_{p}=60~kpc$, and $r_{p}=40~kpc$ respectively.}
    \label{fig:pb_v_by_sigma}
\end{figure}

The above mentioned kinematic decomposition of bulge and disk component requires the prior knowledge of the lesser know dark matter halo distribution. Therefore, we have also shown the evolution of the cold and hot component fractions of the galaxy in terms of observable quantities; mean azimuth velocity ($<v_{\phi}>$) and mean vertical dispersion ($<\sigma_{z}>$). Fig.~\ref{fig:pb_v_by_sigma} shows the time evolution of the ratio of the mean azimuth velocity to mean vertical dispersion ($<v_{\phi}>/<\sigma_{z}>$) within 15~kpc. The left and right panels of the figure represent the 10:1 and 5:1 mass ratio flyby simulations respectively. The left panel clearly shows that the 10:1 flyby models follow the same trend as that of the control isolated model. There is no effect of 10:1 flyby interactions on the evolution of the pseudobulge. But, the 5:1 simulations show a different trends from the control isolated model as can be seen from right panel of the figure. There is more vertical dispersion in the flyby model than the isolated model. This supports the previous results that the flyby interactions with larger mass ratios can heat the pseudobulge more significantly than smaller mass ratios in minor flyby interactions.

\subsection{Mass and Angular Momentum of the Pseudoulge}
\label{sec:pb_mass_angmom}

\begin{figure}
	\includegraphics[width=\columnwidth]{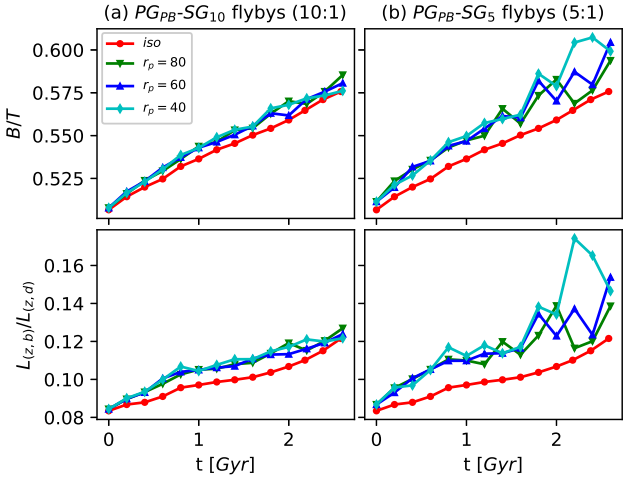}
    \caption{The top and bottom rows show the time evolution of the bulge to total mass ratio ($B/T$) and bulge to disk angular momentum ratio ($L_{(z,b)}$/$L_{(z,d)}$) within 15~kpc respectively. The left and right columns represent the flyby simulations of $1/10$ and $1/5$ mass minor galaxy respectively around the major galaxy with pseudobulge. The red circle solid curves, green down triangle solid curves, blue upper triangle solid curves, and cyan diamond solid curves represent the models $isolated$, $r_{p}=80~kpc$, $r_{p}=60~kpc$, and $r_{p}=40~kpc$ respectively.}
    \label{fig:pb_mass_angmom}
\end{figure}

In the previous subsection, we have discussed the effects of flyby interactions on the kinematic heating and separability of a pseudobulge from the disk. In this section we investigate its effect on the mass and angular momentum of the pseudobulge. To obtain the mass and angular momentum of the bulge using kinematic decomposition, we have used a widely accepted fixed value of $\epsilon=0.7$ as a separator of the bulge and disk particles \citep{Peebles2020}. The particles with $\epsilon<0.7$ are considered as bulge particles and particles with $\epsilon \geq 0.7$ are considered as disk particles.

The time evolution of the bulge to total stellar mass ratio ($B/T$) and the disk to bulge angular momentum ratio ($L_{(z,b)}$/$L_{(z,d)}$) within 15~kpc (the maximum radius for kinematic decomposition) are shown in the top and bottom rows of fig.~\ref{fig:pb_mass_angmom} respectively. The left and right columns represent the flybys of 1/10 and 1/5 mass ratios for a major galaxy with a pseudobulge. Both panels of the left column show the slow increase in the mass and angular momentum of the bulge in flybys relative to the control isolated model. But after the flyby ends, all the models converge to values similar to that of the isolated model. Thus, the 10:1 flyby models do not show any net effect of the interaction on the mass and angular momentum of the pseudobulge. Hence, flyby interactions with relatively low mass satellites show only small changes in mass and angular momentum, and the effect disappears after the satellite has moved away.

On the other hand, the effect of 5:1 flyby interactions are very significant, as can be seen from comparing the adjacent columns of fig.~\ref{fig:pb_mass_angmom}. The mass and angular momentum gained by the pseudobulge remains always larger than the control isolated model. At time t$\sim$2~Gyr, there is approximately 20$\%$ angular momentum gain in flyby interactions relative to the isolated evolution. The trends of mass and angular momentum gain are approximately similar. Therefore, there is a continuous transfer of mass from the disk to the pseudobulge. This is a clear indication of the kinematic heating of the pseudobulge and bulge growth. Hence, relatively massive satellites can help the pseudobulges in flybys to gain mass and angular momentum from the disk. It is possible that the effect of multiple flyby interactions can strongly affect the evolution of psudobulges.

\subsection{Effect of Flyby on the Buckling of the Bar}
\label{sec:bar_buckling}

\begin{figure}
	\includegraphics[width=\columnwidth]{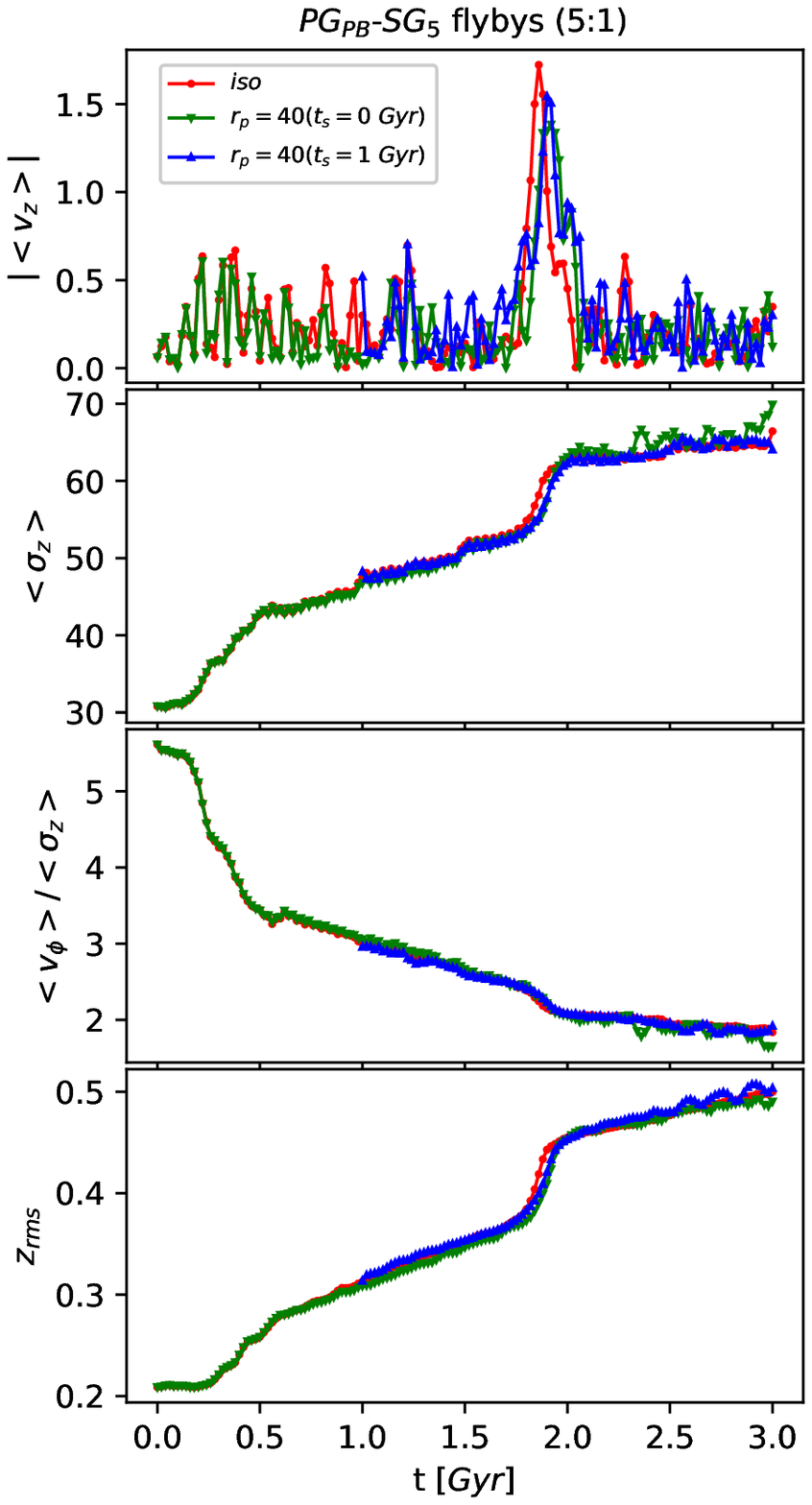}
    \caption{ The time evolution of absolute value of mean vertical velocity($|<v_{z}>|$), mean vertical dispersion ($<\sigma_{z}>$), ratio of the mean azimuth velocity to mean vertical dispersion ($<v_{\phi}>/<\sigma_{z}>$), and root mean square value of vertical coordinates ($z_{rms}$) in the top to bottom panels respectively. The red circle solid curve, green down triangle solid curve, and blue upper triangle solid curve represent the $isolated$, $r_{p}=40~kpc$ starts at $t_{s}=0$ Gyr, and $r_{p}=40~kpc$ starts at $t_{s}=1$ Gyr models respectively.}
    \label{fig:bar_buckling}
\end{figure}

We have previously discussed the response of pseudobulges to flyby interactions with 10:1 and 5:1 mass ratios. The origin of the pseudobulge discussed in this paper is the buckling of the bar. An important question is then how does bar buckling respond to flyby interactions? This is a relevant question as bar buckling modifies the kinematics, rendering the system kinematically hotter. To understand the effect of the flyby interactions on bar buckling we have simulated two 5:1 mass ratio flyby interactions, one starting before the first peak in the bar strength and other starting before the second peak in the bar strength (see fig.~\ref{fig:bar_strength}). The choice of 5:1 mass ratio is made simply because the 10:1 mass ratio does not seem to effect pseudobulges as much as the larger mass ratio.

To quantify the strength of the buckling of the bar, we have adopted some commonly used physical quantities in the literature \citep{Ciambur.etal.2017, Lokas2019}. In  fig.~\ref{fig:bar_buckling}, we have shown the time evolution of absolute value of mean vertical velocity($|<v_{z}>|$), mean vertical dispersion ($<\sigma_{z}>$), ratio of the mean azimuth velocity to mean vertical dispersion ($<v_{\phi}>/<\sigma_{z}>$), and the root mean square value of the vertical positions ($z_{rms}$) in the top to bottom panels respectively. All these quantities are calculated within the 7.5~kpc radius of the galaxies (half of that used in the kinematic decomposition) which is large enough to include the bar region throughout the evolution. The red circle solid curve, green down triangle solid curve, and blue upper triangle solid curve represent the control isolated model, flyby stating at $t_{s}=0$ Gyr from control model, and flyby stating at $t_{s}=1$ Gyr from control model respectively.

From the top panel of the fig.~\ref{fig:bar_buckling}, one can see that all the models show a sharp peak at the same time in the evolution of $|<v_{z}>|$. These peaks occur at the time of the second peak in the bar strength. The peak in the evolution of mean $v_{z}$ signifies the fraction of stars going out of the disk plane. The height of the peak quantifies the magnitude of bar bending out of the disk plane. Though the bending is not strong but it seems that the control isolated model bends the most and flyby interactions reduce the bar bending. The second panel from the top shows the evolution of mean $\sigma_{z}$ which quantifies the the vertical height. All the models show similar evolution of $<\sigma_{z}>$. This result can be further confirmed from the evolution of the other two quantities. The evolution of $<v_{\phi}>/<\sigma_{z}>$, and $z_{rms}$ in the flyby models follows a similar trend as that of the control isolated model. Hence, flyby interactions do not seem to affect the time and strength of the bar buckling in our galaxy model.

\subsection{Effect of Numerical Resolution on Our Results}
\label{sec:numerical_resolution}

\begin{figure}
	\includegraphics[width=\columnwidth]{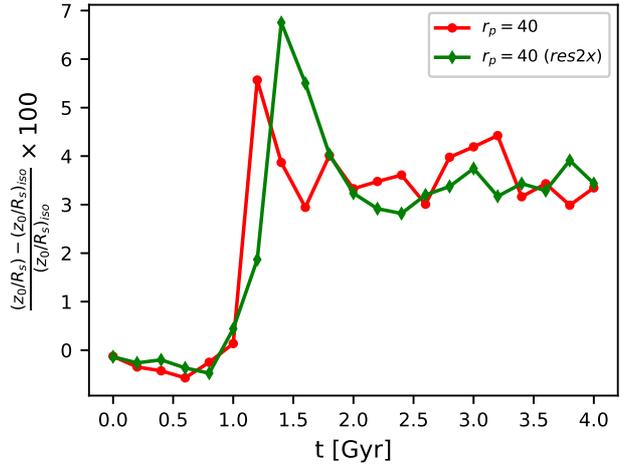}
    \caption{Effect of numerical resolution on the evolution of the percentage change in the disk scale height to disk scale radius ratio in 10:1 flyby model of classical bulge galaxy. The green diamond solid curve represents the simulation of $r_{p}=40~kpc$ model with twice as many particles as red circle solid curve. Both the curves show nearly similar trends.}
    \label{fig:scale_height_res2x}
\end{figure}

All the results of the flyby interactions, discussed in this paper, are relative to the control isolated model. The flyby interactions include secular evolution of the galaxies which can only be removed by comparing with the evolution of an isolated galaxy. On the other hand, numerical simulations always suffer from the discreteness of particle distribution. Some effects of this discreteness on the results of flyby simulations can be removed by subtracting the secular evolution from the evolution in interactions. This is the reason why we compared our results with the corresponding isolated models. We chose the number of particles for our simulation using an analytical study of two body relaxation \citep{Power2003}. We also performed a few high resolution simulations with twice as many particles as given in the Table~\ref{tab:initial_parameters}. We find that the trends as discussed in the results section (section~\ref{sec:results}) are consistent with these high resolution simulation. For comparison, in  fig.~\ref{fig:scale_height_res2x}, we have shown the time evolution of the percentage change in the disk scale height to disk scale radius ratio in the 10:1 flyby model of a classical bulge galaxy with respect to the corresponding isolated model. The green diamond solid curve represents the simulation of $r_{p}=40~kpc$ model with twice as many particles as red circle solid curve. Both curve show more or less same evolutionary trends. This indicates that our results are consistent with high resolution.

\section{Implications}
\label{sec:implication}
In the previous sections, we have presented a qualitative and quantitative study of the effect of flyby interactions on the disks, spiral arms, classical bulges (in unbarred galaxies) and pseudobulges in galaxies. The results of our simulations are very significant for observations of interacting galaxies because in our local universe the rate of flybys is larger at lower redshifts (z$<$2) \citep{Sinha2012A}. In the following paragraphs we discuss the implications of our study for observations of flyby interactions in the local universe.

The gravitational force of a flyby produces strong perturbations in the outer disk regions of the major galaxy. As a result, significant amounts of stellar mass is pulled out from the disk in the form of spiral arms/ tidal tails. This results in the decrease of the disk scale radius ($R_{s}$) and the increase of the disk scale height ($z_{0}$). Although the increase in $z_{0}$ is small, the ratio of disk scale height to disk scale radius ($\frac{z_{0}}{R_{s}}$) shows a significant increase \citep{Reshetnikov1996, Reshetnikov1997}. Since a galaxy experiences multiple flybys during its evolution, the resultant vertical thickening or heating of the host galaxies will be significant and perhaps comparable to that produced by minor mergers.

Flyby interactions stimulate the formation of strong spiral arms in the host galaxies just after the pericenter passage. However, the maximum strength of the spiral arms attained in the flyby does not last. It decays slowly with time and all the flyby models show nearly the same strength in their spiral arms after 4~Gyrs. Our N-body simulations thus show the slowly decaying nature of the spiral arms. This supports the idea that spiral arms are not static density waves as discussed in observations \citep{Masters.2019}. These results will be very helpful in understanding the nature of spiral arms in interacting and isolated galaxies. The spiral arms induced by a closely bound satellite galaxy may survive for a long period of time because a bound satellite will always perturb its host galaxy. The effect of multiple satellite galaxies orbiting around their host on different orbital configurations can give a detailed insight into the origin of the spiral structure.

Strong spiral arms are formed in both the unbarred, classical bulge galaxies and the pseudobulge galaxies. But in contrast to the unbarred, classical bulge galaxies, the flyby induced spiral arms in the pseudobulge host galaxies show oscillations on the top of the time-varying strength of the spiral arms. These oscillations in the spiral arms strength $A_{2}/A_{0}$ could be due to existing small amplitude transients near the resonances associated with the bar \citep{Binney.Tremaine.2008, Wu.etal.2016}, which have been detected in earlier simulations of barred galaxies. The oscillations may be amplified by the energy input due to the flyby interaction. These transient spiral waves may have important implications for star formation and are triggered by the flyby. A good way to test this is by comparing the star formation rates in flybys with host galaxies that have bars and those that do not have bars.

Our overall results of flybys with classical bulges suggests that the inner regions of such galaxies are very stable. Neither the bulge morphology nor the bulge angular momentum changes significantly during flybys. The photometric bulge-disk decomposition of the face-on unbarred, classical bulge galaxies shows a decrease in the mass and sersic index of the classical bulge. This apparent decrease in the sersic index of the classical bulge can be mistakenly considered as evidence for the transformation of the classical bulge into the pseudobulge. So if we take the sersic index as a parameter for distinguishing between classical and pseudobulges, a significant fraction of classical bulges will be mistakenly classified as pseudobulges in face-on galaxies. As a consequence, this classification will provide more weight to the rotation dominated universe and challenge the hierarchical nature of the structure formation \citep{Fisher2011}. So it is clearly important to examine the edge-on decompositions of galaxies whenever possible for sersic index studies of bulges.

Our simulations also show that although the pseudobulges are supported by the ordered motion of the stars, they are very stable against the small perturbations in the outer disk regions of the galaxies. In the 10:1 mass ratio flybys, the pseudobulges do not show any noticeable change except during pericenter passage. There are signatures of dynamical heating, mass growth and angular momentum gain only in the 5:1 mass ratio flybys. Thus only similar mass flybys or large mass ratio minor flybys ($>$1:5) are efficient enough to perturb the evolution of pseudobulges.

Our simulations do not account for the presence of gas in the galaxies. Therefore our findings are valid only for dry or gas poor galaxies. To make a rough estimate of the difference that gas inflow during a flyby would make to the bulge mass, consider the following. Let us assume that the central gas mass fraction of the major galaxy is 5-10$\%$ of the total stellar mass and the star formation efficiency (SFE) is $\sim$10$\%$. During the evolution if the gas is converted to stars, it will contribute 3-6$\%$ of the bulge mass in the center of the galaxy during the flyby. Hence, a gas rich galaxy with a high star formation rate (SFR) may show an increase in bulge mass during such flyby interactions but it will not significantly affect our results.

\section{Conclusions}
\label{sec:conclusions}
We have simulated 10:1 and 5:1 mass ratio flyby interactions of an unbarred, classical bulge galaxy and a barred galaxy with a pseudobulge. We study the effect of the flyby on the major galaxy only. The galaxies are on prograde-prograde orbits, with pericenter distances (distance of closest approach) varying from 40 to 80~kpc. We evolved the flyby for 4~Gyrs and then quantified the evolution of the bulge, disk and spiral arms of the major galaxy using bulge-disk decomposition and Fourier analysis techniques. The main findings of our study are as follows.

1. The disk scale radius ($R_{s}$) of the galaxies decreases in the flyby interactions. The change in the disk scale radius is strongly correlated with the pericenter distance and mass ratio of the galaxies. For a given mass ratio, the close flyby shows the most decrease but for a given pericenter distance, the major flyby (5:1) shows the most decrease. Due to presence of the spiral arms, this change is more pronounced in face-on decomposition than egde-on decomposition. The final disk scale radius seems to converge to the same value for all pericenter distances for a given flyby mass ratio.

2. The absolute disk scale height ($z_{0}$) of the galaxies is marginally affected by the flyby interactions. There is a maximum increase of $\sim 4\%$ in the disk scale height for our set of simulations. However, the ratio of disk scale height to disk scale radius ($z_{0}/R_{s}$) depends significantly on the pericenter distance. The galaxies with smaller pericenter distance show larger changes in this ratio, which is similar to observations of interacting galaxies \citep{Reshetnikov1996,Reshetnikov1997}. But our simulations also show that this is the effect of reducing disk scale length.

3. The minor flyby interactions reduce the disk size but at the same time result in the formation of the grand design spiral arms. The simulations of unbarred, classical bulge galaxies suggests that grand design spiral arms can form in minor flyby interactions when the major galaxy has a concentrated bulge and/or low surface density disk. Both features prevent bar formation in the disk. Hence, grand design spiral galaxies form only in flyby interactions of bar-stable, unbarred galaxies with satellite galaxies.

4. All the models show that the spiral arm strength increases with decreasing pericenter distance passage. However, at the end of the flyby simulation (at 4~Gyrs), all the flyby models show approximately similar spiral arms strength values for a given galaxy mass ratio model, indicating that the final effect does not depend on pericenter distance. The barred, pseudobulge galaxies show small oscillations in the strength of the spiral arms on top of the overall spiral arm strength. The amplitude of these oscillations increases with decreasing pericenter distance. The frequency of these oscillations is approximately independent of the pericenter distances and galaxy mass ratios.

5. The classical bulge of the major galaxy remains generally unaffected by the flyby interactions. In the face-on bulge-disk decomposition, the apparent change in the bulge sersic index is instead due to the steepening of the inner disk surface density and the formation of spiral arms during the flyby. But the edge-on fittings and visualizations of the bulge indicates that the bulge morphology remains unchanged. The kinematic decomposition of the pseudobulges shows some heating and mass growth in the 5:1 simulations. This growth comes at the cost of the disk heating/thickening.

6. There is only a very minute increment in the angular momentum of the classical bulges. This small change is not due to the flyby interactions but due to the rotational motion of the disk particles. The change in the angular momentum of the classical bulges is nearly similar for all the flyby models and isolated models. The evolution of the classical bulges in dry (of gas poor) galaxies remain unaffected of flyby interactions. Pseudobulges show around 20$\%$ gain in the bulge angular momentum after 2~Gyr relative to the secular evolution. But most of it comes from the conversion of disk particles to bulge particles by vertical heating of the disk.

7. The tidal force of the orbiting satellite can induce bar formation in the host galaxy \citep{lang.etal.2014, Lokas.etal.2018}. Tidally induced bars can also buckle with different different strength \citep{Lokas2019}. In our study we find that flyby interactions affect neither the time nor the strength of bar buckling in our galaxy models.


\section*{Acknowledgements}
We thank the anonymous referee for useful comments and suggestions that improved the quality of our paper. We are grateful to high performance computing (hpc) facility `NOVA' at Indian Institute of Astrophysics, Bengaluru, India where we ran all our simulations. We thank Denis Yurin and Volker Springel for providing GalIC code. We thank the authors of the Gadget-2 code that we have used for this study. M.D. acknowledges the support of Science and Engineering Research Board (SERB) MATRICS grant MTR/2020/000266 for this research. S.K. acknowledges the grant MTR/2020/000266 for supporting his visit to IIA during the course of this work. This study also make use of NumPy \citep{numpy2020}, Matplotlib \citep{matplotlib2007}, and Astropy \citep{astropy2018} packages.

\section*{Data Availability}
The data generated in this research will be shared on reasonable request to the corresponding author.




\bibliographystyle{mnras}
\bibliography{Kumar} 




\section{Supplementary Data}
Supplementary data in the form of mp4 videos is available online.



\bsp	
\label{lastpage}
\end{document}